# Exciton-induced magnons carrying orbital angular momentum in $CrI_3$

Martin Pavelka[1]‡, Vishal Shokeen[1]‡, Ruslan Chulkov[1], Soma Dutta[1], David Muradas Belinchón[1], Ulrich Noumbe[1], Mahmoud Abdel-Hafiez[2], M. Venkata Kamalakar[1], Mathias Augustin[3], Vitaliy Goryashko[1], Torstein Hegstad[4], Johan H. Mentink[4], Jamal Berakdar[5,6], Oscar Grånäs[1], Anders Bergman[1,7], Olle Eriksson[1,7], Anna Delin[3,8,9], Hermann A. Dürr[1]*

**Affiliations:**

[1] Department of Physics and Astronomy, Uppsala University, Box 516, 751 20 Uppsala, Sweden.
[2] Department of Applied Physics and Astronomy, University of Sharjah, Sharjah, UAE.
[3] Department of Applied Physics, KTH Royal Institute of Technology, 106 91 Stockholm, Sweden.
[4] Radboud University, Institute for Molecules and Materials (IMM) Heyendaalseweg 135, 6525 AJ Nijmegen, The Netherlands.
[5] Institute of Physics, Martin-Luther-Universität Halle-Wittenberg, 06099 Halle (Saale), Germany.
[6] Halle-Berlin-Regensburg Cluster of Excellence, Center for Chiral Electronics, Germany.
[7] WISE – Wallenberg Initiative Materials Science, Uppsala University, Uppsala, Sweden
[8] Wallenberg Initiative Materials Science, Royal Institute of Technology, Stockholm, Sweden.
[9] Swedish e-Science Research Center (SeRC), KTH Royal Institute of Technology, SE-10044 Stockholm, Sweden.
‡Authors contributed equally
*Corresponding author. E-mail: hermann.durr@physics.uu.se

**Abstract:**

Magnons are collective spin excitations that contain and transport spin angular momentum in magnetic materials. It has been suggested that they can also carry orbital angular momentum in analogy to the electronic motion around the nucleus. We explore the real-space topology of magnon wave-packets emanating from atomic-like excitons in the ferromagnetic insulator $CrI_3$ and demonstrate the existence of orbital angular momentum in such wave-packets. We reveal that orbital angular momentum of magnons is nearly equal to their spin angular momentum and compensates the latter. This illustrates the existence of an unexplored internal angular momentum balance and demonstrates that the magnetization can be quenched without the need of angular momentum exchange with the lattice.

## Introduction

Electronics based on the spin degree of freedom, usually known as spintronics, may lead to developments beyond current silicon-based technologies *(1).* The use of magnons in such technologies could largely avoid the Joule heating from electronic currents in current-day electronics. The discovery of a magnon Hall effect *(2)* represents a crucial milestone to achieve this goal. A magnon wave-packet, picking up a transverse velocity component while moving through the lattice, can be viewed as an object that possesses orbital angular momentum (OAM). In a semiclassical picture, such a twisted wave-packet undergoes rotation around its center *(3, 4).* The number of the rotation periods over a carrier wavelength sets the OAM value which in principle can be very large *(3)*. A key question is tracking the center of this rotation. While this point remains unsolved, it is of practical importance for spin wave engineering via OAM multiplexing *(3)*.

Low-dimensional van der Waals magnetic materials are considered as potential candidates for dissipationless spintronics due to enhanced spin-orbit coupling and a chiral next-nearest-neighbor Dzyaloshinskii-Moriya exchange interaction that can support twisted magnons. The discovery of a chiral exchange interaction induced gap in the magnon band-structure of the 2D ferromagnet $CrI_3$ *(5-7)* (see Fig. 1) seems to support the proposal that such magnons may also carry OAM *(8)*. Yet the size of the OAM associated with a propagating magnon remains unclear *(8)*. First-principles calculations indicate that only a minute amount of it survives, being placed into the crystalline environment of a translational lattice symmetry *(9)*. This is a situation reminiscent of what happens to electronic OAM being quenched by the translational lattice symmetry and the crystalline electric field *(10)*.

Here we show that in $CrI_3$, a prototype 2D-ferromagnetic insulator, the existence of excitons, i.e. bound atomically localized electron-hole pairs *(11),* defines on the atomic scale the center of rotation of magnon OAM, something that is usually lacking for propagating waves. We show that large-amplitude coherent phonons are excited by the generation of $CrI_3$ excitons. This leads, via spin-lattice coupling, to localized spin oscillations around the exciton centers. Finally, for the case where a propagating magnon is excited the measured anomalous spectroscopic signature is interpreted via the generation of a magnon wave-packet carrying OAM of atomic-like size. This has the potential to establish the existence of an angular momentum channel that has not been considered so-far in ultrafast magnetization dynamics *(12-16).*

Figure 1 illustrates the strategy of the paper. The $CrI_3$ lattice lacks inversion symmetry which is even more emphasized when generating an atomic-like exciton (Fig. 1A) *(11, 17, 18).* The existence of a chiral magnetic exchange interaction promoted by the lack of inversion symmetry was shown to induce a gap in the energy dispersion of spin wave (magnon) excitations for certain values of the crystal momentum (Fig. 1B) *(6, 7).* We use an optical excitation to generate excitons *(11, 17),* which in turn excite coherent phonons *(18)* where one of them (the 2.4 THz mode in Fig. 1B) is energetically degenerate with the bottom of the magnon gap in Fig. 1B. In this case propagating magnon modes can be generated via spin-lattice coupling resulting in a chiral spin motion around the excitons in resonance with the 2.4 THz phonon mode (Fig. 1C), as detailed below. We present optical spectroscopy data (see methods and Fig. 1D) that support this scenario and spin dynamics calculations that demonstrate the existence of multi-magnon wave-packets forming around the excitons. These modes are shown to carry OAM of opposite sign to the magnon spin angular momentum. Their formation therefore does not require angular momentum transfer to and from additional angular momentum reservoirs such as the crystal lattice.

**Coherent phonon and spin oscillations following exciton generation**

In this study we utilize the fact that $CrI_3$ displays three bright excitons, A, B and C, at excitation energies of 1.5, 1.9 and 2.2 eV *(19, 20, 11),* respectively, of almost atomic-like spatial dimensions *(11)* that all decay via photoluminescence with roughly 1.2 eV photon energy after microseconds *(17).* The implication of this behavior is that the energy difference between optical excitation and photoluminescence photon energies is stored in $CrI_3$ bosonic excitations, some of which have been observed as coherent phonon oscillations in time-resolved optical spectroscopy *(18, 21-23)*. We observe here a distinctly different behavior for A and B excitons, respectively. In the following, we will show that the stronger bosonic excitation following B-exciton generation leads to propagating magnons surrounding the B-excitons.

Figure 2 displays typical time-resolved optical reflectivity and magnetic rotation measurements when pumping the B exciton. Corresponding data for the A exciton are shown in Fig. S1 (Extended datasets are available in Figs. S2 and S3). Shown in Fig. 2 are the changes in signals of reflectivity, $\frac{\Delta R}{R}$, and magnetic rotation, $\frac{\Delta\theta_M}{\theta_M}$, normalized to the value before arrival of the pump pulse. Data for low fluence (50 μJ/cm$^2$) are shown in Figs. 2A, B and for high fluence (215 μJ/cm$^2$) in Figs. 2C, D. The signal evolution is characterized by an ultrafast signal drop within 200 fs and a subsequent recovery (Figs. 2A, C) or further decay (Figs. B, D) on the ps timescale in agreement with previous reports *(18, 21-23).*

Figure 2 also shows that coherent oscillations are superimposed on the smooth variation of the signal background. We analyzed the amplitude, phase and decay of oscillations in $\frac{\Delta R}{R}$ and $\frac{\Delta\theta_M}{\theta_M}$ for all measured pump fluences (see methods and supplementary text). Selected results are shown as insets in panels (A-D) and the corresponding fast Fourier transform (FFT) amplitudes are displayed in panels (D-G). Extended datasets are shown in Figs. S5 and S6. It is apparent that the phonon oscillations in the $\frac{\Delta R}{R}$ channel are barely visible for exciton A (Fig. S1) while they are pronounced for exciton B (Fig. 2). This agrees with the notion that the higher B-exciton photon energy deposits more energy into bosonic (phonon and spin) excitations compared to the A-exciton *(17).*

The Fourier transforms of the oscillations, shown in Figs. 2E, G, demonstrate the presence of coherent phonon modes at 2.4 and 3.9 THz frequencies due to the Cr-I bond-bending and bond-stretching modes, respectively (see insets in Fig. 2) *(18, 21, 22)*. Determining the amplitudes of these phonon modes for different B-exciton pump fluences results in the linear fluence dependence shown in Fig. S4. We can therefore evaluate the coherent phonon amplitude per exciton by normalizing to the pump fluence. Results are summarized in Fig. 3. The constant phonon amplitude per exciton in Fig. 3A demonstrates that B-excitons are surrounded by a polaronic region containing both 2.4 and 3.9 THz phonon modes that are not influencing one-another even at the highest measured exciton densities, i.e. at the highest pump fluences.

In addition, coherent spin oscillations with identical frequencies as those of the coherent phonon modes are visible in Figs. 2B, D, F, H for B-excitons and Figs. S1B, D, F, H for A-excitons. However, the fluence dependence of the spin oscillations per exciton in Figs. 3B, C is significantly different than that of the phonons in Fig. 3A. Upon A-exciton pumping both 3.9 THz and 2.4 THz spin oscillations display an attenuation with increasing fluence. A similar behavior is observed for the 3.9 THz mode following B-exciton generation. Notably, the 2.4 THz oscillations are absent when B-excitons are generated at low fluences and only appear above a threshold fluence of about 50 μJ/cm$^2$ (Fig. 3B) increasing to amplitudes comparable to that of the 3.9 THz mode at its highest measured fluence.

When excitons are sufficiently close so that their spin oscillations overlap spatially, the resulting interference can be destructive. It is therefore natural to assign to each exciton an effective spin amplitude that decreases monotonically with increasing exciton density. For most cases in Figs. 3B, C this leads to spin oscillations that can be detected optically (see schematics in Fig. 3D). These oscillations are seen to attenuate once exciton neighbors come too close to each other. However, the reverse situation encountered for the 2.4 THz mode after B-exciton formation argues that initially a “hidden” spin oscillation is formed, i.e. one that cannot be seen with the employed optical detection scheme. Spin dynamics calculations, shown below, demonstrate that this corresponds to the formation of large-momentum magnon wave-packets, an interpretation which is further supported on the basis of linear spin-wave theory (see supplementary information).

**Atomistic modeling of phonon-driven coherent spin dynamics**

We first studied the coupling between the observed coherent phonon modes and the magnetic system with so-called frozen phonon calculations (see methods). Displacing the $CrI_3$ lattice according to the Brillouin zone center phonon modes shows that the two phonon modes visible in Fig. 3 couple strongest to the magnetic system. Figure S7A shows that with increasing mode amplitude the effective exchange coupling becomes more strongly modified. However, while the 2.4 THz bond-bending phonon mode increases the exchange coupling, the 3.9 THz phonon mode decreases it. For such a situation, we would expect a characteristic phase relationship to appear in the coherent oscillations of Fig. 2. The results for analysing the phonon and spin oscillation phases vs. fluence are shown in Fig. S7B for B-exciton pumping. There is a clear difference in the phase shifts of 2.4 (red open symbols) and 3.9 THz (blue solid symbols) modes, While the 2.4 THz phonon and spin oscillations modes are in-phase, they are out-of-phase for the 3.9 THz oscillations. We note that this behavior is also apparent from visual inspection of the oscillations shown in Fig. 2. This is exactly as expected from the calculations in Fig. S7A and, thus, indicates that it is the phonon modes that drive the observed spin oscillations localized around the excitons.

With this information we can now address the experimentally “hidden” 2.4 THz spin oscillations at low B-exciton densities. The fact that only large-amplitude B-exciton 2.4 THz phonons display this phenomenon (the 2.4 THz mode is clearly present for A-excitons in Fig. S1F) indicates that only they can drive large-momentum magnons as visualized in Fig. 1B. We use atomistic spin dynamics calculations to address this formation of traveling magnons generated via exciton formation. The wavelength of such magnons would be below the detection limit of the optical spectroscopy tools employed here as for such frequencies the magnon momenta are close to the Brillouin zone boundary (see Fig. 1B) *(6, 7)*.

Figure 4 shows typical results (see methods) of propagating magnon wave-packets generated by the 2.4 THz phonon-induced modulation of the exchange interaction around an exciton (positioned at the origin). Figure. 4A shows a snapshot of the spin motion at 100 fs after exciton formation. It reflects the torque exerted by the chiral exchange interaction on the next-nearest-neighbor spins around the exciton (see Fig. 1C and methods). The initial spin orientation clearly displays a chirality around the central exciton. This chirality is transferred onto magnon wave-packets that propagate outwards from the exciton with a phase difference that depends on the azimuthal angle as can be seen in the longer-time snapshot of Fig. 4B and the supplementary movies S2, S3.

**Orbital angular momentum (OAM) analysis of topological magnon wave-packets**

The most interesting feature in Fig. 4 is the formation of spiral-like phase fronts indicated in the panel B. While the colored regions in Fig. 4B correspond to spin orientations along the positive/negative x-direction, the white regions represent the corresponding spin orientations along the y-direction (see extended dataset in Fig. S10). Different temporal snapshots demonstrate that the spiral rigidly rotate in a clockwise direction (see Fig. S10). Such a spiral pattern can only form if the individual magnons propagating along different crystalline directions in the wave packet have a non-zero phase relationship relative to each other. Moreover, the spiral structure performs one complete clockwise rotation while the atomic spins undergo one counterclockwise rotation (see the supplementary movies S1-3). We will show in the following that this corresponds to an antiparallel alignment of spin and orbital angular momentum components for the magnon wave-packets.

Figure 4C shows the calculated temporal evolution of the total spin modification, $\Delta S_z$, and induced orbital, $L_z$, angular momentum of the propagating magnon wave-packet. It is important to note that the atomic magnetic moment, $\boldsymbol{M}$, precession (see inset of Fig. 4C) is related to the change of spin angular momentum as $\Delta M_z = -\frac{g\,e}{2m}\Delta S_z$ with $g$ being the g-factor, $e$ the absolute value of the electron charge and $m$ the electron mass. The magnon spin angular momentum can be obtained directly from the magnetic moment reduction along the z-direction integrated over all atoms (blue curve in Fig. 4C). Since each magnon lowers the total spin angular momentum of the system by $1\hbar$ ($\hbar$ is Planck's constant divided by $2\pi$) the total change in spin angular momentum integrated over all atoms is proportional to the total number of excited magnons, $N_{total}$ with $\Delta S_z = N_{total}\,\hbar$.

The situation is different for the total magnon OAM, $L_z$, which is obtained from the characteristic phase changes of the magnetic moment precessions for adjacent atoms, e.g. the spiral features in Fig. 4B. We evaluated $L_z$ (red curve in Fig. 4C) from the OAM of one magnon, $L_{z,magnon}$, (see methods) as $L_z = L_{z,magnon}\,N_{total}$. We note that the OAM per magnon, $L_{z,magnon}$, is related to the spatial phase relation between neighboring precessing magnetic moments which is in principle can be different in size to the atomic spin precession contributing to $\Delta S_{z,magnon}$.

Figure 4C demonstrates the existence of OAM, $L_z(t)$, in the spin motion around a central exciton. For the first 1 ps the temporal variation of $L_z(t)$ and $\Delta S_z(t)$ differ since it takes time before the spiral precession pattern in Fig. 4B is fully formed (see methods and supplementary movie S2). However, at longer times $L_{z,magnon}(t)$ becomes roughly constant and the total amount of OAM increases due to the formation of magnons near the central excitons and their subsequent damping during the outwards propagation (see methods). This shows that the magnon wave-packets in Fig. 4 surrounding excitons in $CrI_3$ carry zero total angular momentum (i.e. $L_z + \Delta S_z = 0$). Consequently, no angular momentum needs to be transferred from a reservoir to generate such magnon wave-packets.

Figure 4D displays the radial spin and orbital angular momentum extend around the central exciton at two different times (4 ps, open symbols and full lines and 5 ps, solid symbols and dotted lines). The insets of Fig. 4D demonstrate that the spiral magnon wave-packet contains magnons that propagate outwards from the exciton. We estimate their average group velocity to be approximately 10 Å/ps from the respective wavefront shifts. These propagating magnons are likely those that are close in energy-momentum space, E(k), to ground-state magnons (see methods and Fig. S9). The main part of the spiral magnon wave-packet is, however, spatially localized around the exciton with a spatial extend of about 20Å (Fig. 4D). This is composed of the non-equilibrium magnon content of the wave-packet within the gap of the ground-state E(k) magnon dispersions (see methods and Fig. S9).

While the above results show that for the magnon wave-packets generated by excitons in $CrI_3$ the total linear and angular momentum is zero, it is still necessary for all to receive their energy from the exciting 2.4 THz phonon modes. We have studied experimentally the damping of the coherent phonon amplitudes of 3.9 and 2.4 THz modes (see Fig. S7C) and obtained damping times of $\tau_{2.4} = 24.0 \pm 3.0$ ps and $\tau_{3.9} = 47 \pm 12$ ps, respectively. The 3.9 THz phonon mode excites localized spin oscillations that are damped on longer times that the measured data range (see extended datasets in Figs. S5, 6). The 2.4 THz phonon mode is more strongly damped even in the absence of a localized spin precession (Fig. 1F). It is the energy degeneracy of the 2.4 THz phonon mode with magnons that carry OAM (see Fig. 1B) that can facilitate stronger energy transfer. This process is blocked for the 3.9 THz phonon mode as no magnons with OAM exist at this energy.

We note that the spiraling dynamic magnetization $M(r,t)$ in Fig. 4B has a well-defined rotation which enables future sensing and possibly steering via electrical means: In absence of free charges and charge currents, this follows directly from $\nabla \times \left(\frac{\dot{B}}{\mu_0} - \dot{M}\right) = \epsilon_0 \epsilon_r \ddot{E}$, leading to $\Delta E + \frac{\epsilon_r}{c^2}\ddot{E} = \partial_t(\nabla \times M)$. Here, $\mu_0, \epsilon_0, c$ are the permeability, permittivity and speed of light in vacuum. $\epsilon_r$ is $CrI_3$ dielectric constant. Clearly, $\nabla \times M$ acts as a source for the electric field $E$ with $-\dot{B} = \nabla \times E$.

We would like to stress that previous studies considered orbital angular momentum primarily as an intrinsic property of propagating Bloch magnons in reciprocal space *(9),* obtaining typically very small values. In our case, the exciton that generates the localized phonon modes provides a natural real-space center of rotation and hence, a large OAM for the excited magnons. This provides an additional angular momentum reservoir that can compensate spin angular momentum changes, and it is therefore not necessary to involve chiral phonons *(24-26)* to generate magnon excitations.

**Acknowledgments:** The Wallenberg Initiative Materials Science for Sustainability (WISE) funded by the Knut and Alice Wallenberg Foundation is acknowledged.

**Funding:**

Swedish Research Council (VR) grant 2022-02881 (H.A.D.)

Swedish Research Council (VR) grant (O.E.)

Swedish Research Council (VR) grant 2024-04986 (A.D.)

Swedish Research Council (VR) grants 2021-05932 and 2024-05531 (M.V.K.)

Carl Trygger Foundation grant CTS 23:2560 (H.A.D.)

Knut and Alice Wallenberg Foundation (KAW) grant 2022.0108 (H.A.D., A.D., O.E)

Knut and Alice Wallenberg Foundation (KAW) grant 2022.0079 (M.V.K.)

Knut and Alice Wallenberg Foundation Scholar grant (O.E.)

European Research Council (ERC) synergy grant FASTCORR (O.E.), eSSENC (O.E.), 3D-MAGiC (J.H.M.)

Dutch Research Council (NWO), VIDI project no. 223.157 CHASEMAG (J.H.M)

European Research Council (ERC) grant 101002772 (M.V.K.)

Deutsche Forschungsgemeinschaft (DFG) grants 429194455 and 465098690 (J.B.)

Some of the computations/data handling were enabled by resources provided by the National Academic Infrastructure for Supercomputing in Sweden (NAISS), partially funded by the Swedish Research Council through grant agreement no. 2022-06725.

**Author contributions:**

Conceptualization: M.P., V.S., H.A.D.

Methodology: M.P., V.S., R.C., H.A.D., M.A., A.B., M.A.H., D.M.B., U.N., M.V.K., V.G., T.H., J.H.M., J.B., O.G., O.E., A.D.

Investigation: M.P., V.S., R.C., H.A.D., M.A., A.B., M.A.H, D.M.B., U.N., M.V.K., V.G., T.H., J.H.M., J.B., O.G., O.E., A.D.

Visualization: M.P., S.D., H.A.D., A.B.

Funding acquisition: H.A.D., M.V.K., J.H.M., O.E., A.D., J.B.

Supervision: H.A.D., M.V.K., J.H.M., A.B., O.G., O.E., A.D.

Writing – original draft: H.A.D.

Writing – review & editing: M.P., V.S., R.C., S.D., D.M.B., U.N., M.A.H., M.V.K., M.A., V.G., T.H., J.H.M., J.B., O.G., A.B., O.E., A.D., H.A.D.

**Competing interests:** Authors declare that they have no competing interests.

**Data and materials availability:** All data needed to evaluate the conclusions in the paper are present in the paper and/or the Supplementary Materials. Derived data supporting the findings of this study is available at https://doi.org/10.5281/zenodo.21682490

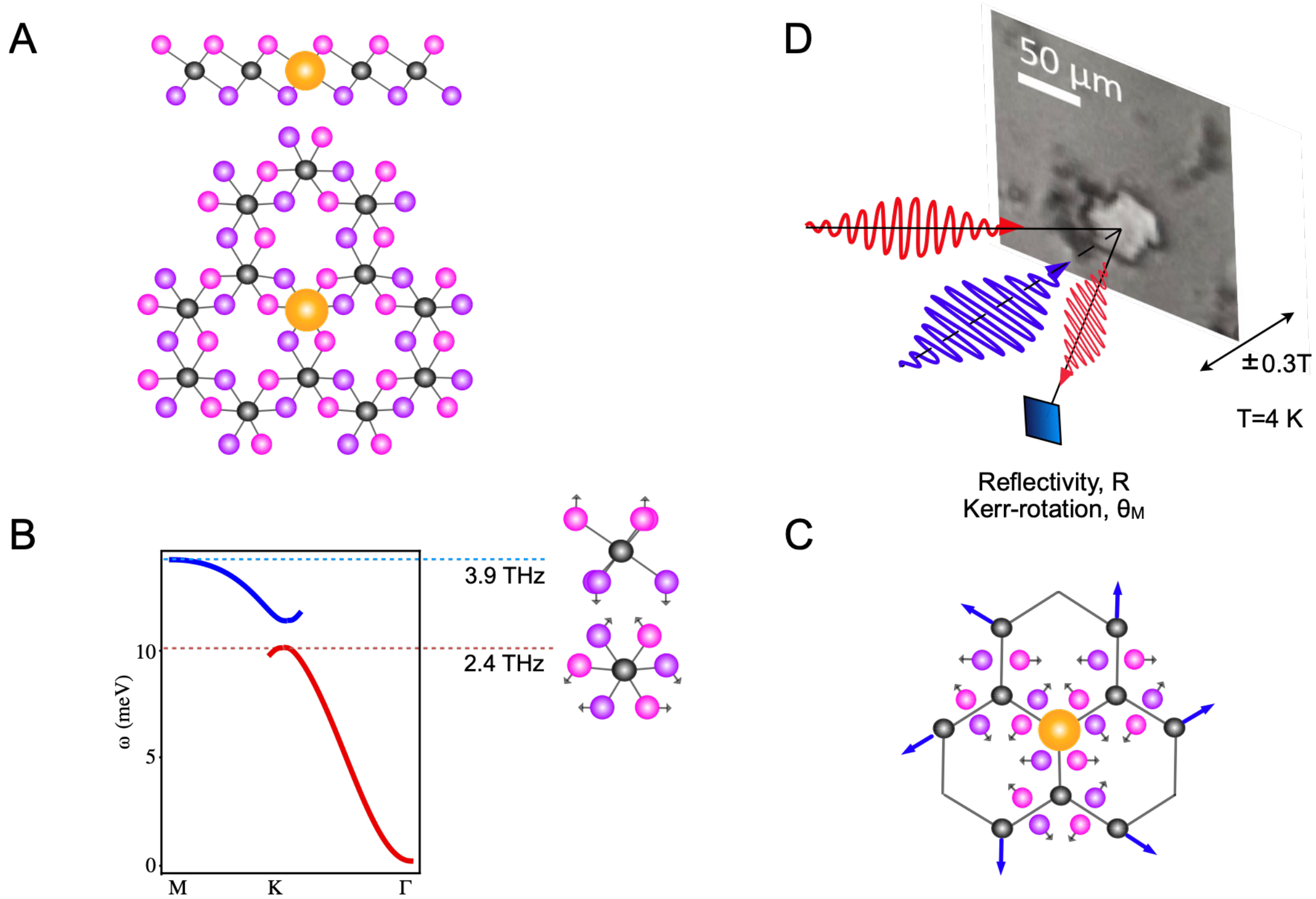

**Fig. 1. $CrI_3$ lattice, magnons, chiral spin torque and optical pump-probe measurements.** (**A**) Real-space side (top panel) and top (bottom panel) view of a $CrI_3$ with Cr ions marked as black and I ions as color small circles. The large yellow circles indicate excitons. (**B**) Exciton formation leads to the generation of coherent phonons of 3.9 THz and 2.4 THz frequencies and with displacements of the I ligands as schematically shown by arrows. These phonon frequencies are close to optical (blue line) and acoustic (red line) magnon energies near the M- and K-points, respectively, of the $CrI_3$ Brillouin zone *(7)* as shown on the left of the panel (see methods). (**C**) Exciton induced spin torque due to the next-nearest-neighbor chiral exchange interaction (see methods) leads to a chiral magnetic precession pattern (blue arrows) that can couple to K-point magnons. (**D**) Optical pump-probe measurements on 50 μm size $CrI_3$ flakes are described in the methods section.

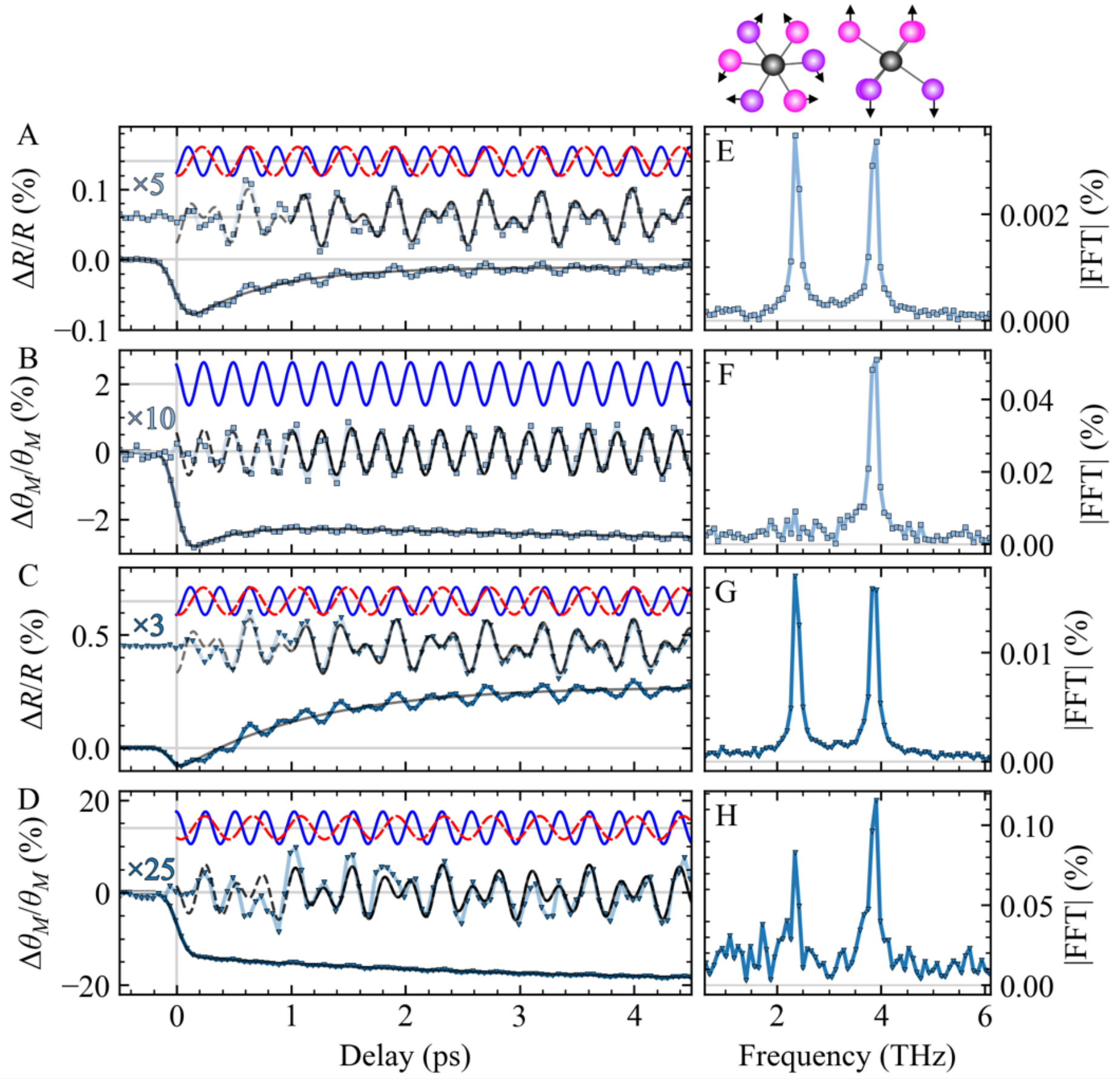

**Fig. 2. Optical B-exciton pump – C-exciton probe measurements.** Optical measurements near the C-exciton resonance (hν = 2.4 eV) while pumping the B-exciton resonance (hν = 1.9 eV). (**A**, **C**) Reflectivity and (**B**, **D**) Kerr-rotation measurements vs. pump-probe time delay at pump fluences of 50 μJ/cm$^2$ (**A**, **B**) and 215 μJ/cm$^2$ (**C**, **D**). Measured data are shown as blue symbols and lines, with smooth background fits (solid black lines, see methods) and the background-subtracted signals with the indicated enlargement factors. The dashed black lines represent fits of the coherent oscillation measurements with the contributions of the 2.4 THz and 3.9 THz phonon modes (see insets above (**D**)) shown as red and blue solid lines, respectively. In (**E**, **F**, **G**, **H**) we display absolute amplitudes of fast Fourier transforms (FFT) vs. frequency for the background-subtracted data in (**A**, **B**, **C**, **D**), respectively. In inset the phonon modes of the two excitations are shown (see text). Similar measurements at the A-exciton resonance are shown in Fig. S1 for comparison.

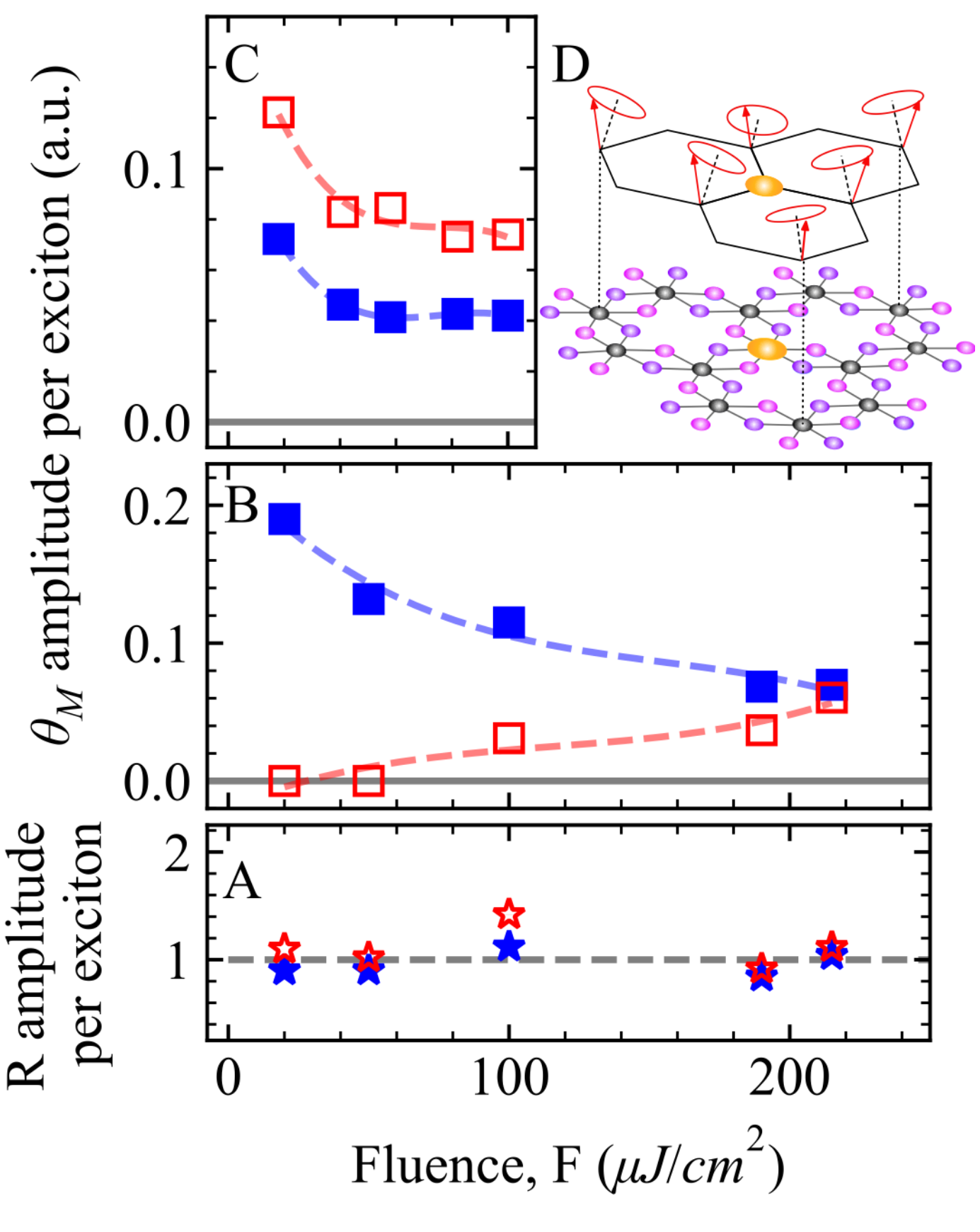

**Fig. 3. Exciton dependent phonon and spin oscillations.** (**A**) Amplitudes of the 2.4 THz (open red symbols) and 3.9 THz (solid blue symbols) phonon oscillations normalized to the number of generated B excitons (see extended dataset in Fig. S3a). (**B**) Amplitudes of the 2.4 THz (open red symbols) and 3.9 THz (solid blue symbols) spin oscillations normalized to the number of generated B excitons, i. e. divided by the pump fluence (see extended dataset in Fig. S3b). (**C**) Same as (**B**) but for the A excitons (see extended dataset in Fig. S2b). Dashed lines in (**B**, **C**) represent guides to the eye. (**D**) Illustration of the spin oscillation (red arrows) around an exciton (yellow circle).

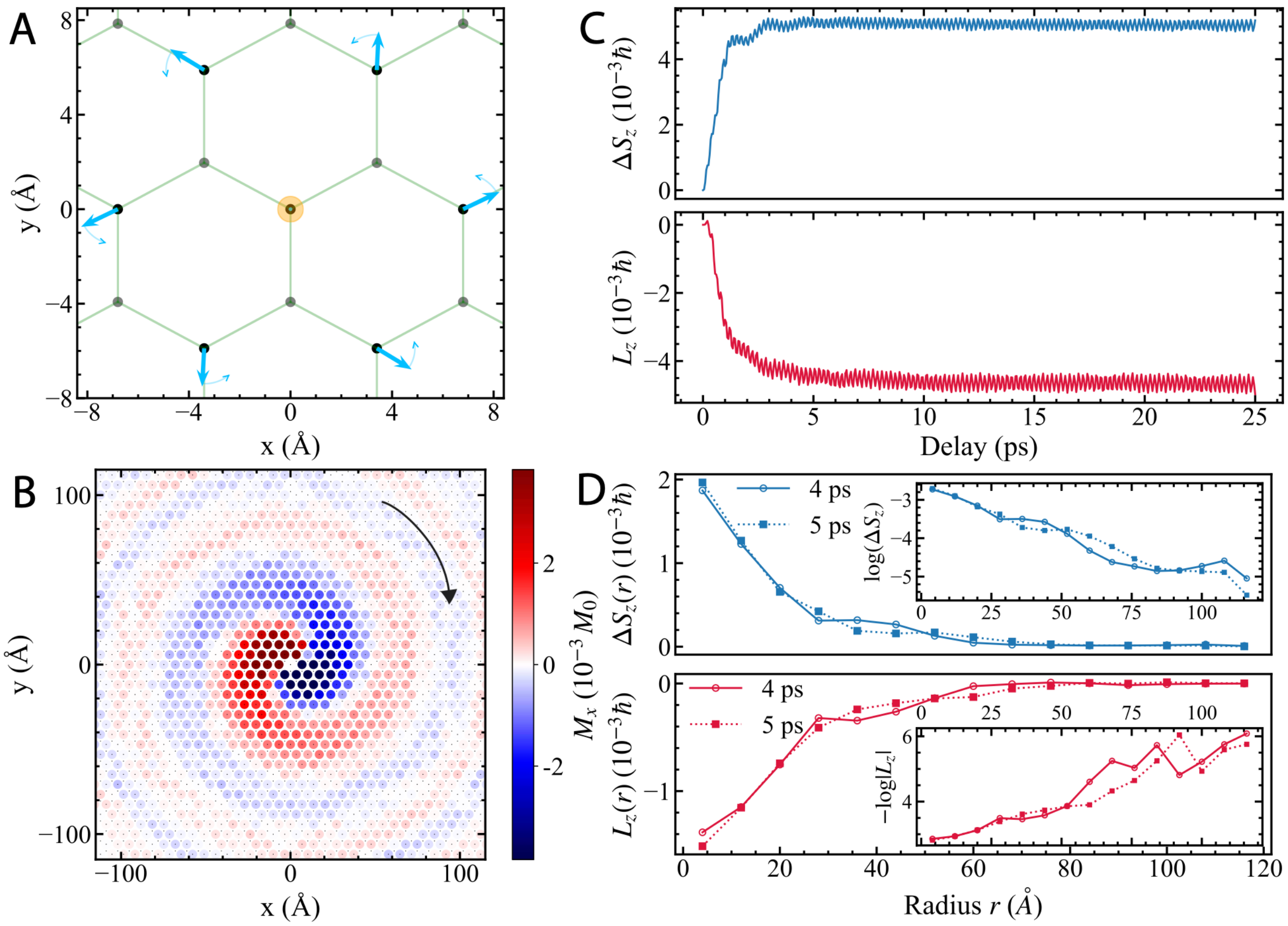

**Fig. 4. Calculated exciton-generated spin motion for one magnetic sublattice.** (**A**) Atomistic spin dynamics calculations where the exchange interactions are modulated via the 2.4 THz phonon mode (see Fig. 1C and methods) result in the displayed atomic magnetizations (blue arrows) after 100 fs. Subsequently, (the magnetic moments precess counterclockwise (blue curved arrows). This represents the magnetic A-sublattice connected to the exciton (yellow circle) via chiral exchange interactions (see Fig. 1C and Fig. S8). Subsequently, also the magnetic B-sublattice, connected to the exciton via linear magnetic exchange interactions, start to precess (see methods and Movie 1). (**B**) Atomic $M_x$ magnetization components for the A-sublattice at 10.0 ps display a spiral-like pattern that precess clockwise (see Movie 2 and black curved arrow). (**C**) displays the temporal evolution of the change of spin, $\Delta S_z$, and orbital, $L_z$, angular momentum components integrated over all atoms in the calculated slab. The inset illustrates the reduction of the magnetic moment component, $\Delta M_z$, when initially static moments, $M_0$, start to precess. (**D**) Temporal snapshot at 4 ps (open circles and dotted lines) and 5 ps (solid squares and solid lines) of the radial distribution of $\Delta S_z$ and $L_z$, vs. radial distance, $r$, from the exciton displays features due to localized and propagating magnons. The inset shows log-plots to emphasize the propagating magnon feature. Extended datasets of the spin motion in (**A**) and (**B**) for both magnetic sublattices are available as supplementary movies and Fig. S10.

## Materials and Methods

Sample preparation and characterization

$CrI_3$ crystals are prepared using were synthesized using the chemical vapor transport (CVT) method and characterized with x-ray diffraction and magnetometry *(27-29)*. The resulting $CrI_3$ crystals exhibit a plate-like morphology, with lateral dimensions reaching several millimeters. The crystals were stored in an inert-gas atmosphere glove box to prevent degradation *(30)*. Samples were freshly cleaved inside the glove box prior to further characterization. For the time-resolved magneto-optical Kerr effect (MOKE) experiment *(23)*, samples were mounted in a LHe cryostat inside the glove box. The bulk like $CrI_3$ flakes were exfoliated using tape and transferred onto a 300 nm thick $SiO_2$ film on a Si substrate. The samples inside the cryostat were then transferred to the MOKE experiment.

The size of the $CrI_3$ flakes used in the experiment was approximately 50 μm as imaged in tr-MOKE set-up itself (see Fig. 1D). This also allows us to monitor the physical degradation of the sample as $CrI_3$ is notoriously sensitive to the environmental conditions and exhibit low laser damage threshold *(27)*. The magnetic signature in samples is confirmed by measuring the polarization rotation signal for two field directions in the absence of pump beam below $T_c$. The saturation magnetic field for bulk $CrI_3$ is ~0.2 T *(27-30)*. The static magnetic polarization rotation for our samples is ~25 mrad measured at a photon energy of 2.4 eV.

The laser fluence used in the experiment is constrained by the signal to noise ratio for the lowest fluence and damage threshold for the largest fluence. We kept well below about 20% of the damage threshold ensuring the long-term stability of the samples. We note that $CrI_3$ samples used here gave consistent results as measured across several months as they were kept under an inert Ar atmosphere during preparation and subsequently high vacuum (~3 $10^{-6}$ mbar) conditions during the experiment *(23)*.

Experiment

A femtosecond laser (CARBIDE -CB3, Light Conversion) generated the 190 fs long (FWHM) pulses at the pulse repetition rate of 12 KHz with the photon energy of 1.2 eV at the fundamental harmonics. Part of the fundamental laser beam was frequency doubled to 2.4 eV in a BBO crystal and later used as an optical probe. The other beam part was directed at an optical parametric amplifier (ORPHEUS -ONE-HP, Light Conversion) to generate 125±15 fs long pulses within the 1.5 - 1.9 eV photon energy range. The OPA pulses were used as an optical pump. The overall time resolution determined by the pump-probe pulse convolution was estimated to be ≈200 fs. The linearly p-polarized probe beam and pump beam of the same polarization were directed on a sample at angles of ~22° and ~0° wrt the sample normal, respectively. The 50 mm and 100 mm focal length lenses provided the probe and pump beam spots equal to 15±5 and 38±5 μm (FWHM) at the sample position. The peak pump fluence was measured to be 215±60 $\mu J/cm^2$. The polar-MOKE reflection geometry with magnetic field strength of ±0.3 T was provided by two pairs of permanent Nd magnets. A special translational holder allowed the magnet pairs to be swapped. The probe beam reflected by the sample was recollimated back by a 50 cm focal length lens. A polarizing beam splitter (the Wollaston prism) was used for spatial separating the p- and s- beam polarizations. The balance probe signal detection was provided by a half-waveplate and two identical Si photodetectors (DET36A2, Thorlabs). Short-pass filters before the photodetectors (FESH0550, Thorlabs) guaranteed suppression of the optical pump crosstalk. Amplitudes of the photodiode signals were digitized by a two-channel locking amplifier (SE1022D, Saluki Technology) at the modulation frequency of 2.09 kHz that was maintained by a mechanical chopper (MC2000B, Thorlabs).

The pump-probe optical delay was controlled with an absolute accuracy of ±2 fs by a motorized stage (DDSA04, Thorlabs). Each set of measurements was performed for two opposite orientations of the magnetic field with a delay step equal to ≈50 fs. An average sum signal from both the photodetectors was then processed to find the relative change in the sample reflectivity. The magnitude of the Kerr rotation was determined by subtracting the imbalance signals of the two detectors.

Frozen phonon calculations

The ground-state geometry and properties were calculated using the Vienna Ab-Initio Simulation Package (VASP) *(31)* version 5.4, a plane-wave-based implementation of density functional theory. For the theoretical analysis, we base our DFT calculations on the $CrI_3$ monolayer with the following computational details. We used the local spin density approximation (LSDA) for the exchange correlation functional where the core electrons were represented via scalar relativistic Projector Augmented Wave pseudopotentials. The plane-wave basis set had an energy cutoff of 350 eV and the momentum space was sampled by a Γ-centered 12 × 12 ×1 k-point mesh. The atomic coordinates and the cell size were optimized until the interatomic forces were less than $10^{-6}$ eV/Å. The first-principles calculation of phonon modes was performed using the finite-displacement method with VASP and Phonopy version 2.12 *(32)* using a 4x4x1 supercell and 3x3x1 k-point mesh.

In order to verify that we capture the relevant physics with our monolayer geometry and choice of calculation options, we also performed similar calculations on bulk $CrI_3$. There we used the generalized gradient approximation of Perdew, Burke, and Ernzerhof for solids *(33)* for the exchange-correlation functional and van der Waals corrections were introduced through the zero-damping DFT-D3 method of Grimme *(34)*. The plane-wave basis set for the bulk calculations had an energy cutoff of 500 eV, and the momentum space was sampled by a Γ-centered 6 × 11 ×11 k-point mesh. The atomic coordinates and cell size were also here optimized until the interatomic forces were smaller than $1 \times 10^{-6}$ eV/Å. The first-principles calculation of the bulk phonon modes was performed using the finite displacements method using VASP *(31)* and phonopy version 2.12 *(32)*. The results are consistent at 2×2×2 and 3×3×3 supercells, indicating that phonons are converged with respect to cell size.

The exchange interactions have been calculated for the monolayer using the methodology outlined in Ref. *(35)* Using energy differences between chosen magnetic states, we calculate the effective isotropic Heisenberg exchange and chiral Dzyaloshinskii-Moriya exchange interactions between Cr ion sites *i* and *j* with unit vectors $\hat{e}_i$ and $\hat{e}_j$ of the form

$$\hat{H} = -\sum_{i\neq j} J_{ij}\hat{e}_i\,\hat{e}_j - \sum_{i\neq j}\overline{D}_{ij}\cdot(\hat{e}_i\times\hat{e}_j) \qquad \text{(M1)}$$

including all interactions within a five lattice site radius. In eq. (M1) $J_{ij}$ and $\overline{D}_{ij}$ represent the linear and chiral exchange interactions, respectively. Furthermore, exchange interactions, $J_{ij}(u_k)$ and $D_{ij}(u_k)$, have also been calculated for the displacement pattern of normal phonon mode, $k$, with displacement amplitude $u_k$. This way the effect of lattice distortions on the exchange interactions is obtained.

Based on the above methodology we calculate the coupling between the spin-system and the polaronic lattice distortions forming around the excitons. We performed calculations for all Raman active phonon modes. The results for modes that show the strongest spin-phonon coupling are displayed in Fig. S7A. The $u_k$ dependence for the next nearest neighbour isotropic, $J_2^\alpha(u_k)$, and chiral exchange, $D^\alpha(u_k)$, constants for bonds along $\alpha = x, y, z$ coordinates for the $u_{2.4THz}$ mode is shown in Fig. S8.

Estimates of the effective exchange interactions from the bulk calculations showed a similar magnon-phonon coupling between the two different geometries. Thus, we found that both bulk and monolayer calculations essentially capture the same physical picture for the relevant electronic, magnetic and structural degrees of freedom.

Atomistic spin dynamics simulations

The calculated exchange interactions (see Figs. S7A and 8) are used as input for modelling the magnetic excitations in the system. The atomistic spin dynamics (ASD) simulations have been performed using the UppASD code *(36)* and we refer to *(37)* for details regarding the general implementation.

The ASD simulations include all magnetic interactions calculated for the monolayer albeit with a renormalization in order to ensure that the top of the acoustic magnon band at the K point coincides with the frequency of the 2.4 THz phonon mode. This renormalization does not change the underlying physics of the system but just facilitates a comparison between theory and experiment.

In order to mimic the effect of the exiton on the spin dynamics, we model the system as follows. We assume that the exciton excites the 2.4 THz phonon mode and investigate how this in turn can drive magnetic excitations. The phonon calculations, as outlined above, provide the time-dependent lattice distortions around the exciton, assumed to be localized on a single Cr atom *(11)*. Combined with the calculated distance dependence of the exchange interactions, $J_{ij}(u_k)$, the lattice distortions can be interpeted as dynamically modifying the local exchange interactions. We confine ourselves to phonon amplitudes for which the modified exchange interactions display a linear variation (see Fig. S7A) and write

$$J_{ij}(t) = J_{ij}{}^{0} + \Delta J_{ij}\, sin(\omega t)$$
$$\overline{D}_{ij}(t) = \overline{D}_{ij}{}^{0} + \Delta\overline{D}_{ij}\, sin(\omega t) \qquad \text{(M2)}$$

where $\omega$ is the phonon frequency. We consider the exciton to be localized on a single site $i$ and the sign and magnitude of $\Delta J_{ij}$ and $\Delta\overline{D}_{ij}$ are then determined from symmetry analysis of the phonon modes combined with the calculated variation of the exchange interactions with lattice displacemnts. We found that exchange pairs $ij$ up to second-nearest neighbours are the most important to consider as being dynamically modified. Results for the effective bulk exchange, $J_{bulk}$, are shown in Fig. S7A and for the isotropic next-nearest neighbor exchange, $J_2^{iso}(u_k)$, in Fig. S8A. For next-nearest neighbors also chiral exchange plays a role *(6, 7)*. Figures S8 B-D displays the chiral exchange coupling paprameters $D_2^\alpha(u_k)$ obtained for the second term in eq. (M1) for the exciton-Cr bonds indicated in the inset.

When pumping the 2.4 THz magnon using the full symmetry of the exchange interactions in Fig. S8 we obtain magnon wavepackets localized at the K-point as shown in Fig. S9B. We note that a reduction of the phonon-magnon coupling symmetry, i.e. switching off the $D_2^{\alpha}(u_k)$ term, in Fig. S9C leads to negligible magnon excitations for both magnetic sublattices.

To analyze the dynamical properties of the excited magnon, we introduce the complex magnon magnetization $\psi(r,t) = N^{-1}\left[M_x(r,t) + i\, M_y(r,t)\right]$ where $M_x$ and $M_y$ are the site-resolved transverse components of the magnetization in relation to the initial magnetic ground state $M_0$ oriented along the positive z-direction with the normalization $N = \left|\sum_r M_x(r,t) + i\, M_y(r,t)\right|$. From the definition of the orbital angular momentum (OAM) operator we obtain $\hat{L} = \bar{r} \times \hat{p} = -i\hbar(\bar{r} \times \nabla)$. The position $r$ is measured with respect to the position of the exciton. The z-component of the OAM operator $\hat{L}_z = -i\hbar(x\partial_y - y\partial_x)$ yields the intrinsic orbital angular momentum carried by the magnons. Applying these operators to the magnon wavefunction $\psi(r,t)$ we obtain the orbital angular momentum density per magnon, $L_{z,magnon}(r,t)$, and the integrated orbital angular momentum per magnon, $L_{z,magnon}(t)$ shown in Fig. 4.
The dissipation of energy and subsequent damping of the excited magnons depend on the intrinsic Gilbert damping of the system. To test the effect of this damping, we examined the behaviour when varying the Gilbert damping parameter α. Since the exchange pumping from the central exciton is persistent throughout our simulations, the results were rather insensitive to the choice of damping for reasonable values of $\alpha$. Varying $\alpha$ within the interval $0.0001 < \alpha < 0.01$ resulted in very similar dynamical behaviour of the excited magnetic texture. The major difference is that the extent of the excited region grows with smaller $\alpha$, which is expected since the energy dissipation is then reduced. The effect is also visible for zero damping albeit too small values of $\alpha$ makes the analysis of the excited spin texture difficult since the lack of energy dissipation from the spin system results in that the excited region grows larger than the simulation cell.

We illustrate the effects of phonon-magnon coupling in eq. (M2) by evaluating the magnon spectral function $S(q,\omega)$ in Fig. S9 along the M −K −Γ high-symmetry directions for the $CrI_3$ monolayer. S(q, ω), corresponds to the time and space Fourier transform amplitude obtained for the $CrI_3$ monolayer via atomistic simulations shown as movie M2. Clearly visible are the nonequilibrium magnons at the driving frequency of 2.4 THz (horizontal intensity line at 10.4 meV). These are the ones that carry orbital angular momentum (see Fig. 4). The part of these magnons in resonance with the ground-state magnon dispersions (curved lines in Fig. S9) at the K-point represent the propagating part of the magnons seen in Fig. 4D. The part of the driven magnons with wavevectors away from the ground-state magnon branches contribute to magnons localized near the exciton site. Note that ground-state magnon dispersion (curved lines) in Fig. S9 are in good agreement with the dispersion relations obtained with linear spin wave theory *(35)*.

Linear spin wave theory (LSWT)

To support experimental and atomistic simulation results, linear spin wave theory was performed. We use the spin model described in *(7)* and also their parameters:

$$H_0 = J_1 \sum_{\langle ij\rangle} \mathbf{S}_i \cdot \mathbf{S}_j + J_2 \sum_{\langle\langle ij\rangle\rangle} \mathbf{S}_i \cdot \mathbf{S}_j + J_3 \sum_{\langle\langle\langle ij\rangle\rangle\rangle} \mathbf{S}_i \cdot \mathbf{S}_j + \sum_{\langle\langle ij\rangle\rangle} \mathbf{D}_{ij} \cdot (\mathbf{S}_i \times \mathbf{S}_j) + K \sum_i (S_i^z)^2$$

where $S_i$ is the spin operator at lattice site $i$ on a honeycomb bipartite lattice. $J_1 = -2.11$ meV is the nearest-neighbor (NN) exchange interaction, $J_2 = -0.11$ meV is the next-nearest-neighbor (NNN) exchange interaction, $J_3$=0.1 meV is the next-next-nearest-neighbor (NNNN) exchange and $K = -0.123$ meV is the easy plane anisotropy. The third term is an out-of-plane NNN DMI interaction with $\mathbf{D}_{ij} = \nu_{ij} D_z \mathbf{z}$ where $\nu_{ij} = +1(-1)$ for clockwise (counterclockwise) hopping along the hexagons. $D^z = 0.09$ meV.

To be consistent with atomistic simulations, where the magnetization is initialized along $+\mathbf{z}$, we here initialize the *spin* along $-\mathbf{z}$. To diagonalize the Hamiltonian we start by using the Holstein-Primakoff (HP) transformation with $S_i^z = -S + c_i^\dagger c_i$, $S_i^+ = \sqrt{2S} c_i^\dagger$, and $S_i^- = \sqrt{2S} c_i$ where $c_i = a_i(b_i)$ on sublattice A(B). $S$ is the spin value and $S_i^\pm = S_i^x \pm i\, S_i^y$. For the description of spin waves, it is convenient to work in reciprocal space using the Fourier transform given by $c_i = \sqrt{2/N} \sum_k e^{ikr_i} c_k$ where $N$ is the number of lattice sites. Ignoring constant terms, we obtain $H_0 = \sum_k \Psi_k^\dagger H_0(k) \Psi_k$ where $\Psi_k^\dagger = [a_k, b_k]$ and

$$H_0 = \begin{bmatrix} h_k^0 + h_k^z & h_k^{xy} \\ (h_k^{xy})^* & h_k^0 - h_k^z \end{bmatrix},$$

$$h_k^{xy} = J_1 S \gamma_k + J_3 S q_k, \qquad h_k^0 = -3J_1 S - 6J_2 S - 2KS + 2J_2 S p_k, \qquad h_k^z = -2D_z S \rho_k.$$

In the above expressions, $\gamma_k = \sum_\delta e^{ik\delta}$ with $\delta$ being the NN lattice vectors, $p_k = \sum_\zeta \cos(k \cdot \zeta)$ and $\rho_k = \sum_\zeta \sin(k \cdot \zeta)$, where $\zeta$ are the three NNN vectors representing counterclockwise jumping. Lastly, $q_k = \sum_\xi e^{ik\xi}$, where $\xi$ are the NNNN lattice vectors. Next, we introduce the Bogoliubov transformation $\Psi_k = U_k \Phi_k$ with

$$U_k = \begin{bmatrix} u_k e^{i\varphi_k} & -v_k e^{i\varphi_k} \\ v_k & u_k \end{bmatrix}, \qquad \Phi_k = \begin{bmatrix} \alpha_k \\ \beta_k \end{bmatrix}, \qquad u_k = \sqrt{\frac{\epsilon_k + h_k^z}{2\epsilon_k}}, \qquad v_k = -\sqrt{\frac{\epsilon_k - h_k^z}{2\epsilon_k}}$$

where $\varphi_k$ is defined by $\gamma_k = |\gamma_k| e^{i\varphi_k}$. Using the Bogoliubov transformation, we diagonalize the Hamiltonian

$$H_0 = \sum_k \omega_k^\alpha\, \alpha_k^\dagger \alpha_k + \omega_k^\beta \beta_k^\dagger \beta_k,$$

where

$$\omega_k^\alpha = h_k^0 + \epsilon_k, \qquad \omega_k^\beta = h_k^0 - \epsilon_k, \qquad \epsilon_k = \sqrt{|h_k^{xy}|^2 + (h_k^z)^2}.$$

To later calculate orbital angular momentum (OAM), we need the in-plane components of the spins. Inspired by *(3),* we define

$$\psi_i = \langle S_i^x \rangle + i \langle S_i^y \rangle = \langle S_i^+ \rangle = \sqrt{2S} \langle c_i^\dagger \rangle,$$

and using the same Bogoliubov transformation, we can write the expectation values of the HP bosons as

$$\langle a_i^\dagger\rangle = \sqrt{\frac{2}{N}}\sum_k e^{-ikr_i}\, e^{-i\varphi_k}(u_k\langle\alpha_k^\dagger\rangle - v_k\langle\beta_k^\dagger\rangle), \qquad \langle b_i^\dagger\rangle = \sqrt{\frac{2}{N}}\sum_k e^{-ikr_i}\,(v_k\langle\alpha_k^\dagger\rangle + u_k\langle\beta_k^\dagger\rangle)$$

Next, we calculate the dynamics of $\psi_i$ under an oscillatory perturbation of DMI:

$$\delta H = f(t)\sum_{\langle\langle i,j\rangle\rangle} \Gamma_{i,j}\,\mathbf{D}_{i,j}\cdot(\mathbf{S}_i\times\mathbf{S}_j)$$

where $f(t) = \sin(\Omega t)\Theta(t)$ and $\mathbf{D}_{i,j} = [D^x\cos\theta_{ij} \quad -D^y\sin\theta_{ij} \quad 0]$, where $\theta_{ij}$ is defined by the NNN lattice vector $\mathbf{r}_{ij}/|\mathbf{r}_{ij}| = [\cos\theta_{ij} \quad \sin\theta_{ij}]$, and here we have used $D^x = D^y = 0.2$ meV. $\Gamma_{i,j}$ is the perturbation strength over bond $i,j$. In our case, it will be 1 for perturbed bonds, and 0 for the rest. More specifically, we consider one lattice site at the origin on the A sublattice. Following a similar process as described for diagonalization, we get

$$\delta H = f(t)\sum_k [A_k\alpha_k + B_k\beta_k + h.c], \qquad A_k = \sqrt{\frac{S^3}{N}}G_k u_k e^{i\varphi_k}, \qquad B_k = -\sqrt{\frac{S^3}{N}}G_k v_k e^{i\varphi_k}$$

$$G_k = -2i\sum_\zeta D_\zeta \sin(k\cdot\zeta), \qquad D_\zeta = -iD^x\cos\theta_\zeta - D^y\sin\theta_\zeta.$$

We can now calculate the expectation values of the magnon operators in linear response

$$\begin{aligned}\langle\alpha_k^\dagger\rangle &= -i\int_{-\infty}^{t} d\,t'\langle[a_{k,I}^\dagger(t),\delta H_I(t')]\rangle = iA_k e^{i\omega_k^\alpha t}\int_{-\infty}^{t} d\,t' f(t')e^{-i\omega_k^\alpha t'} \\ &= iA_k\frac{-1}{2}\left[\frac{e^{i\Omega t}-e^{i\omega_k^\alpha t}}{\Omega-\omega_k^\alpha} + \frac{e^{-i\Omega t}-e^{i\omega_k^\alpha t}}{\Omega+\omega_k^\alpha}\right]\end{aligned}$$

and equivalent for $\langle\beta_k^\dagger\rangle$. The subscript $I$ refers to the interaction picture; $O_I = e^{iH_0t}Oe^{-iH_0t}$. To model decay in the system, we let $\omega_k \to \omega_k + i\epsilon$. We are also interested in the magnon density distribution, and to calculate this we use second order response:

$$\langle n_k^\alpha\rangle = \langle\alpha_k^\dagger\alpha_k\rangle = -\int_{-\infty}^{t} d\,t_1\int_{-\infty}^{t_1} d\,t_2\langle[\delta H_I(t_2),[\delta H_I(t_1), n_{k,I}^\alpha]]\rangle = |\langle\alpha_k^\dagger\rangle|^2$$

By driving the system with the magnon frequency at the bottom of the gap at the $K_1$ *(7)*, $\Omega = \omega_{K_1}^\beta$, we as expected excite magnons around the $K_1$ point. Figure S11 shows the magnon density distribution on top of the dispersion relations.

To calculate the OAM of the system, we write the linear momentum of a magnetic moment as *(8)*

$$\mathbf{p}_i = \frac{1}{2S}[\langle S_i^y\rangle\nabla\langle S_i^x\rangle - \langle S_i^x\rangle\nabla\langle S_i^y\rangle] = \frac{1}{2S}\mathrm{Im}\{\psi_i\nabla\psi_i^*\}$$

We are working on a discrete lattice, thus, we approximate the spatial derivative by $\nabla\psi_i = \frac{2}{3}\sum_\delta \delta\,(\psi_{i+\delta} - \psi_i)$ where the factor $2/3$ comes from the geometry of the honeycomb lattice. This lets us calculate the total out-of-plane orbital angular momentum of the system:

$$L_z = \sum_i [\,\mathbf{r}_i \times \mathbf{p}_i]_z$$

Fig. S12 shows the time evolution of the OAM and the total change of magnetization

$$\langle \Delta S^z \rangle = \sum_i \langle\, c_i^\dagger c_i \rangle = \sum_k \langle\, \alpha_k^\dagger \alpha_k \rangle + \langle \beta_k^\dagger \beta_k \rangle.$$

It is also important to mention that when modeling the magnon behavior under the perturbation, the unperturbed Hamiltonian can be simplified. In fact, to achieve the results shown here, $H_0$ can be reduced to only nearest neighbor exchange. In addition, since the linear momentum is proportional to $\psi^2$, the overall sign of the perturbation does not matter. As mentioned earlier, we only perturb next-nearest neighbor DMI of one lattice site in one sublattice. The total OAM is independent of which sublattice you choose, but the OAM distribution is rotated equivalent to the required rotation to go between the two sublattices. Fig. S13 shows the OAM distribution.

**Supplementary Text**

Fitting the average temporal evolution

We fit the measured data of the two observables $\Delta R/R\ (t)$ and $\Delta\theta_M/\theta_M\ (t)$ using:

$$F(t) = A(t, \tau_{pulse})\left[F_{0,exc} + F_{0,heat}(1 - e^{-t/\tau_{heat}}) + F_{0,decay}(1 - e^{-t/\tau_{decay}})\right], \qquad \text{(SE1)}$$

where $A(t, \tau_{pulse})$ is a Heaviside step function at t=0 convoluted by a Gaussian with full-width half maximum $\tau_{pulse}$. Furthermore, the size of the initial, instantaneous response is characterized by its amplitude $F_{0,exc}$, which corresponds to the number of excitons created by the pulse. It is followed by an ultrafast heating process with amplitude $F_{0,heat}$ and time constant $\tau_{heat}$, eventually extending into a subsequent decay process with amplitude $F_{0,decay}$ and time-constant $\tau_{decay}$. This last term is happening on a time-scale longer *(22)* than our experimental time range, hence, in most of the datasets, is fixed to 1000 ps. The fitting results are presented graphically in Figures S1a,b and S2a,b, and the fitting parameters in Table S1. The aforementioned fixed long time constant is indicated by a start symbol* because it is not varied during the fitting procedure. This procedure was based on fitting using the least squares method, minimizing the sum of the squares of the residuals, with errors determined as one-standard-deviation uncertainties, calculated as the square roots of the diagonal elements of the covariance matrix.

Fitting phonon and spin oscillations

We further fit the residual of the above procedure using two decaying sine modes:

$$G(t) = G_{LF}\sin(2\pi\omega_{LF}t + \varphi_{LF})e^{-t/\tau_{LF}} + G_{HF}\sin(2\pi\omega_{HF}t + \varphi_{HF})e^{-t/\tau_{HF}}\ , \qquad \text{(SE2)}$$

where the frequencies are fixed as $\omega_{LF}$= 2.375 THz and $\omega_{HF}$= 3.865 THz, having amplitudes $G_{LF}$ and $G_{HF}$, starting phase $\varphi_{LF}$ and $\varphi_{HF}$ and decay times $\tau_{LF}$ and $\tau_{HF}$. The fitting results are presented graphically in Figures S5, S6a and S6b, and the fitting parameters in Table S2. The corresponding oscillation amplitudes for exciton B reflectivity are presented in Figure S7.

We further elaborate the crucial result of decay times of reflectivity of Exciton B, by comparing three fitting strategies: no decay, one decay and two decay constants. We use these assumptions, fit, and extract the standard deviations σ and $R^2$ calculated as 1 minus the ratio of the residual sum of squares to the total sum of squares. The option with two decay constants always results in a better fit, the results are shown in Table S3.

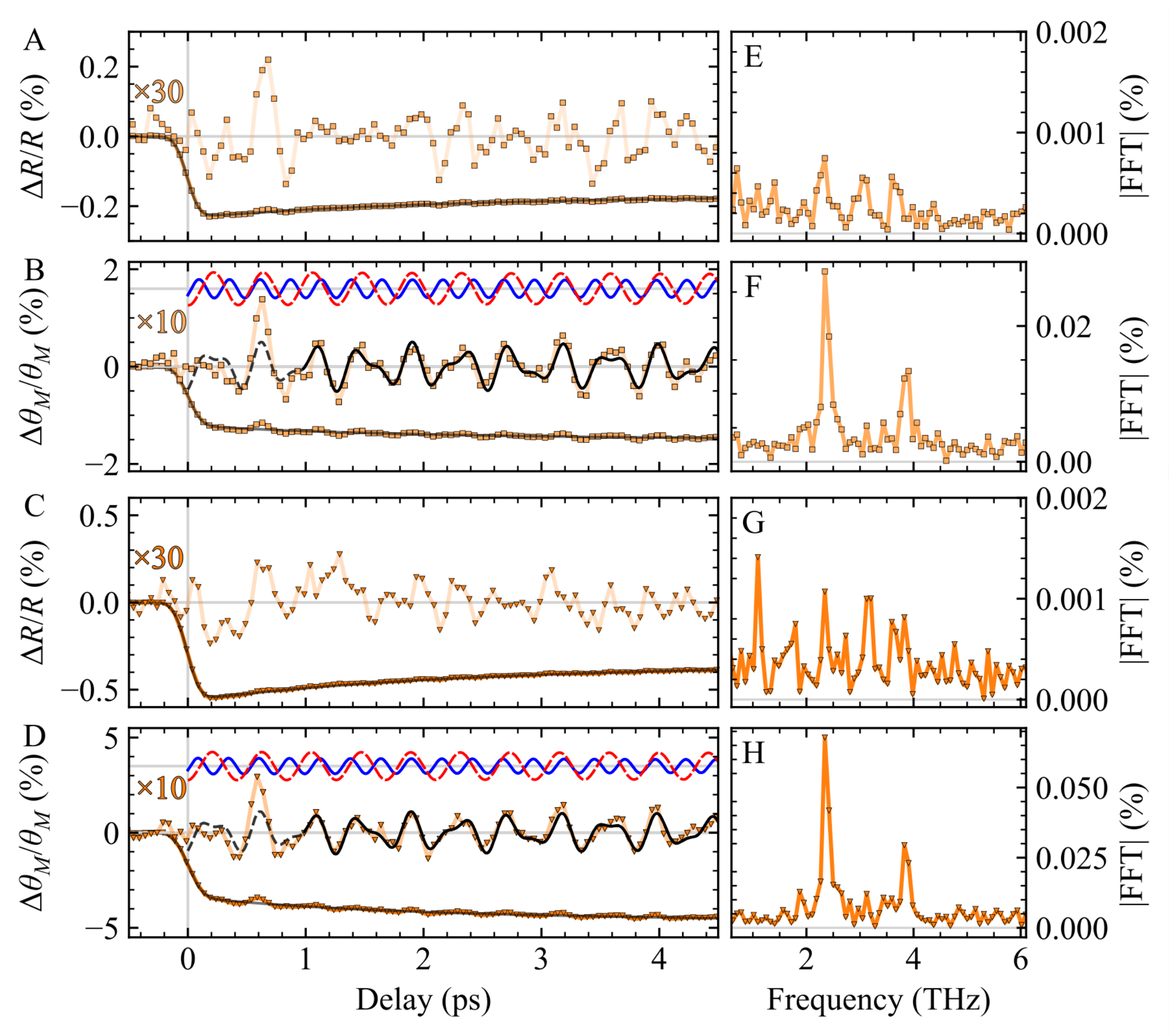

**Fig. S1. Optical A-exciton pump – C-exciton probe measurements.** Optical measurements near the C-exciton resonance (hν = 2.4 eV) while pumping the A-exciton resonance (hν = 1.5 eV). (**A**, **C**) Reflectivity and (**B**, **D**) Kerr-rotation measurements vs. pump-probe time delay at pump fluences of 41 μJ/cm$^2$ (**A**, **B**) and 100 μJ/cm$^2$ (**C**, **D**). Shown are the measured data as orange symbols and lines, smooth background fits (solid black lines, see methods) and the background-subtracted signals with the indicated enlargement factors. The dashed black lines represent fits of the coherent oscillation measurements with the contributions of the 2.4 THz and 3.9 THz phonon modes (see insets in (**D**)) shown as red and blue solid lines, respectively. (**E**, **F**, **G**, **H**) display absolute amplitudes fast Fourier transforms (FFT) vs. frequency for the background-subtracted data in (**A**, **B**, **C**, **D**), respectively.

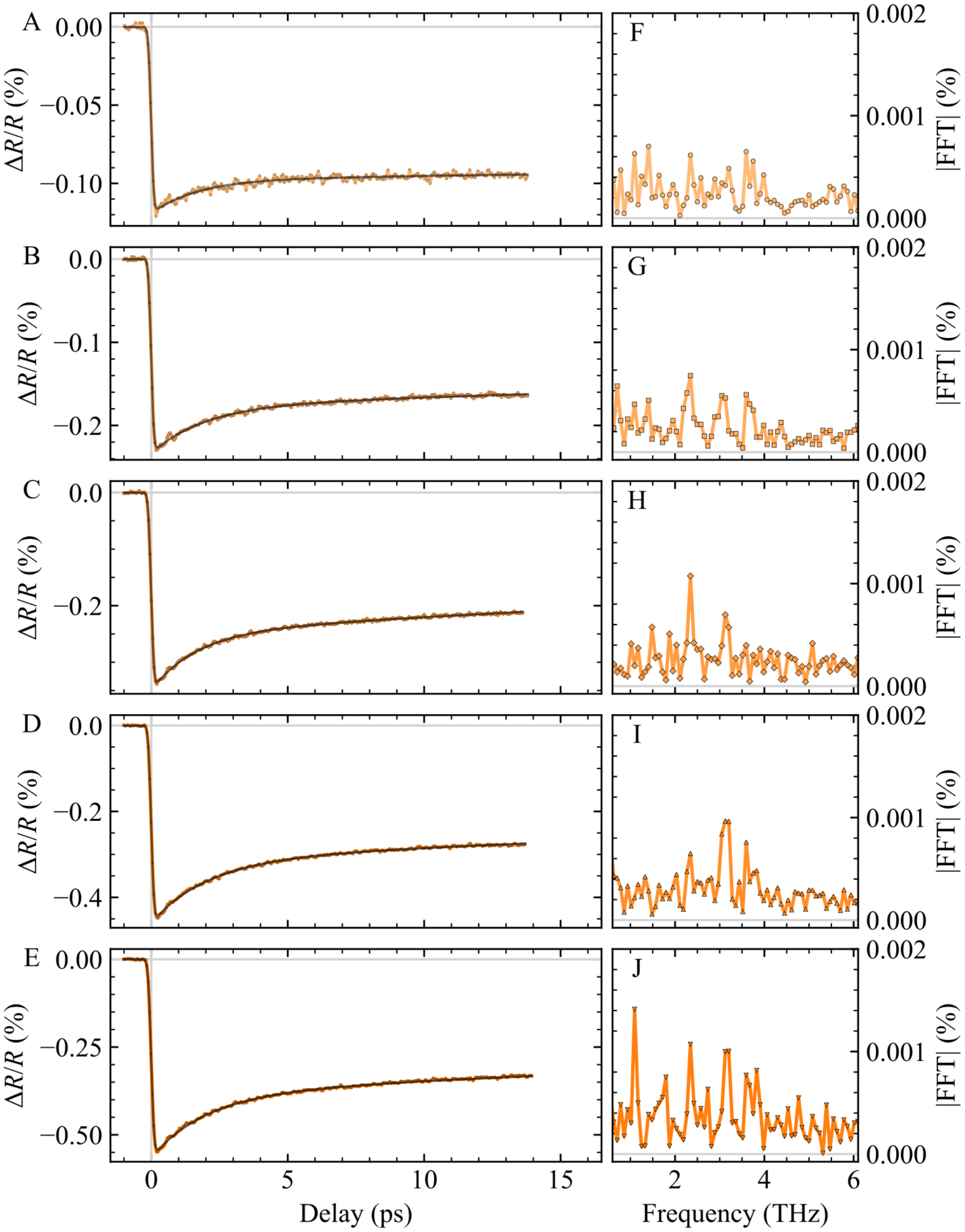

**Fig. S2a. Extended dataset for the A exciton, reflectivity.** Panels A-E show the measured data with the average temporal evolution fit for varying fluences (18, 41, 58, 82 and 100 μJ/cm$^2$, respectively). These measurements were recorded concurrently with those shown in Fig. S2b. Panels F-J show the fast Fourier transforms of the fit residuals. The fit parameters for the solid line are summarized in Table S1.

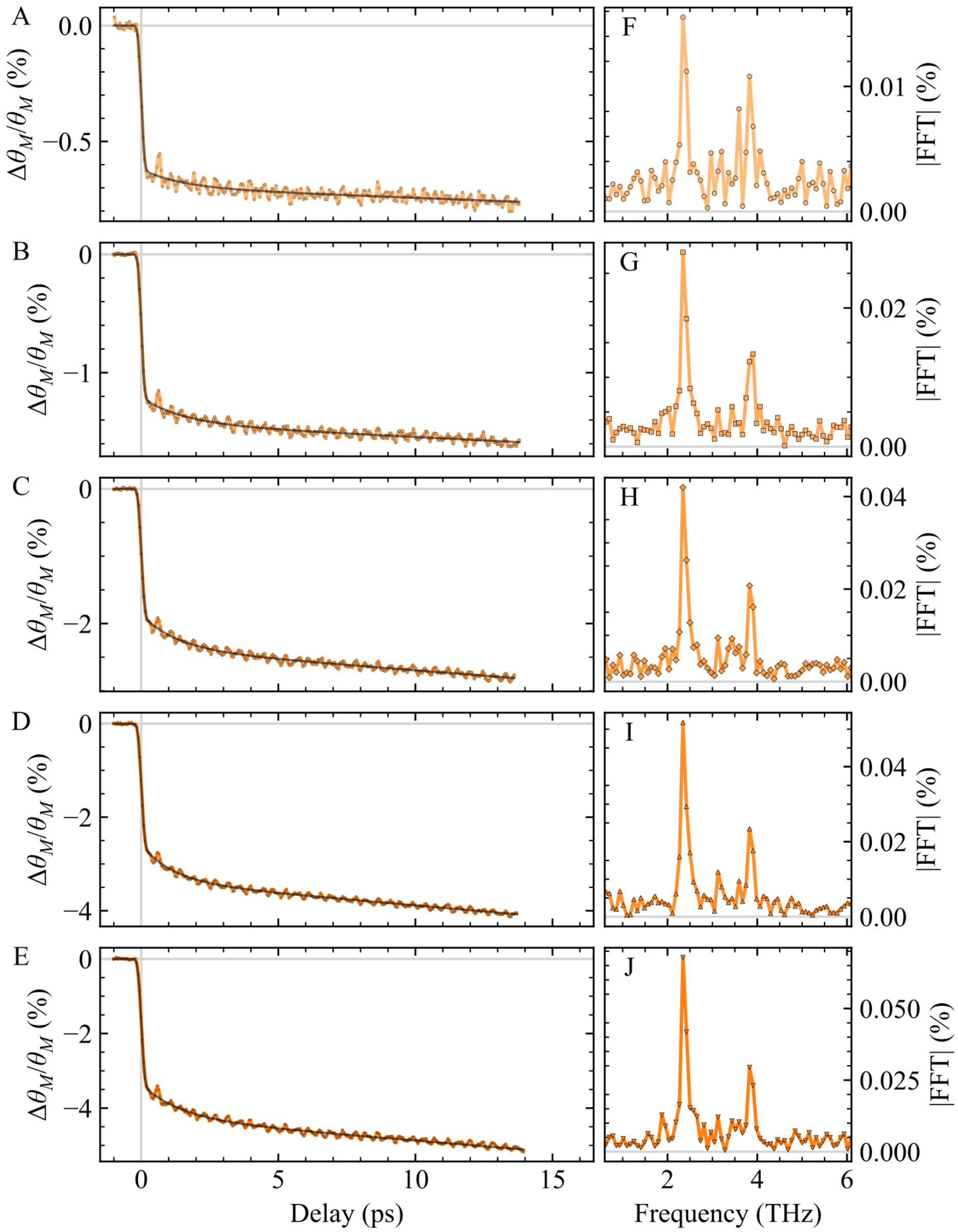

**Fig. S2b**. Extended dataset for the A exciton, magnetic rotation. Panels A-E show the measured data with the average temporal evolution fit for varying fluences (18, 41, 58, 82 and 100 μJ/cm$^2$, respectively). These measurements were recorded concurrently with those shown in Fig. S2a. Panels F-J show the fast Fourier transforms of the fit residuals. The fit parameters for the solid line are summarized in Table S1.

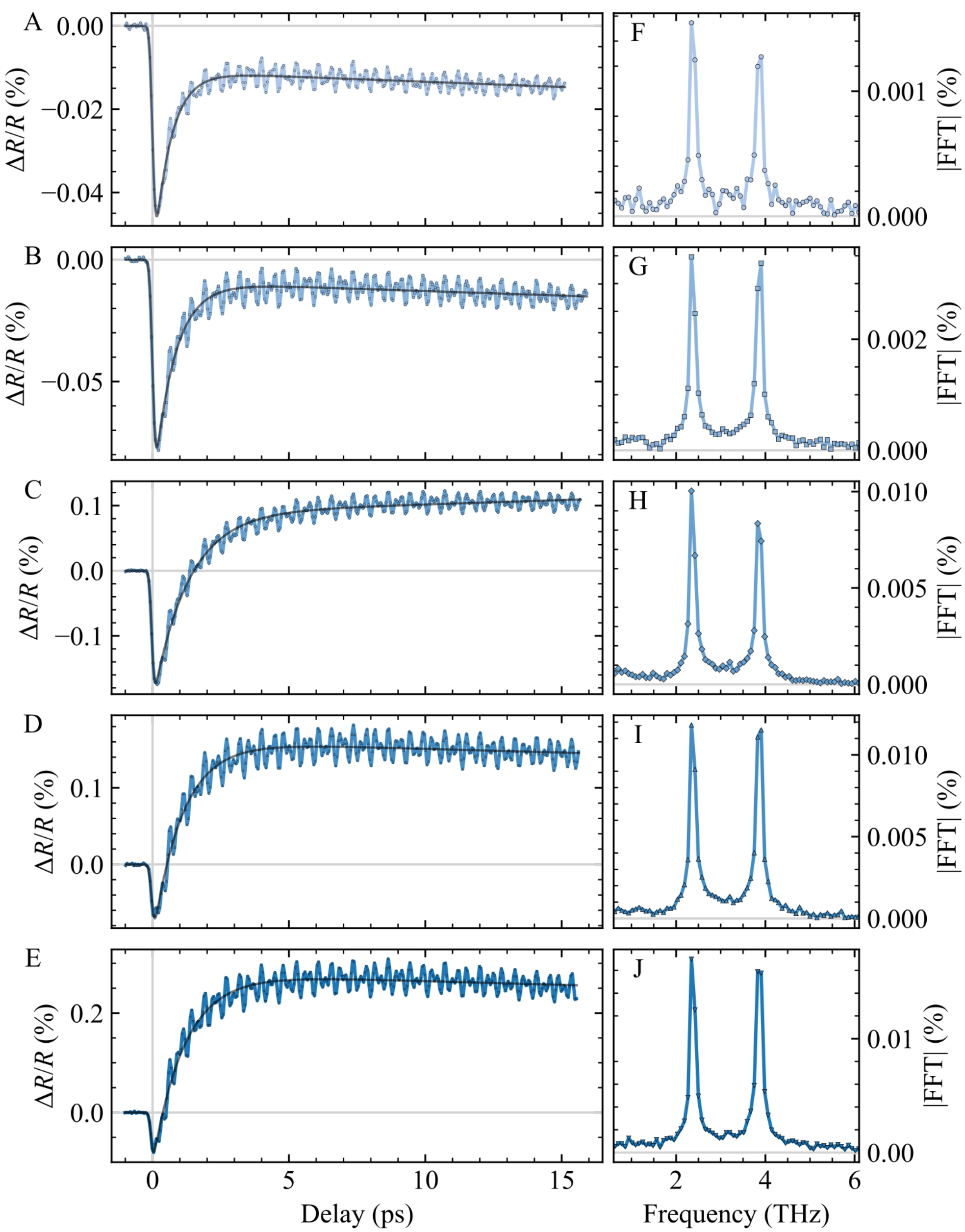

**Fig. S3a. Extended dataset for the B exciton, reflectivity.** Panels A-E show the measured data with the average temporal evolution fit for varying fluences (20, 50, 100, 190 and 215 μJ/cm$^2$, respectively). These measurements were recorded concurrently with those shown in Fig. S3b. Panels F-J show the fast Fourier transforms of the fit residuals. The fit parameters for the solid line are summarized in Table S1.

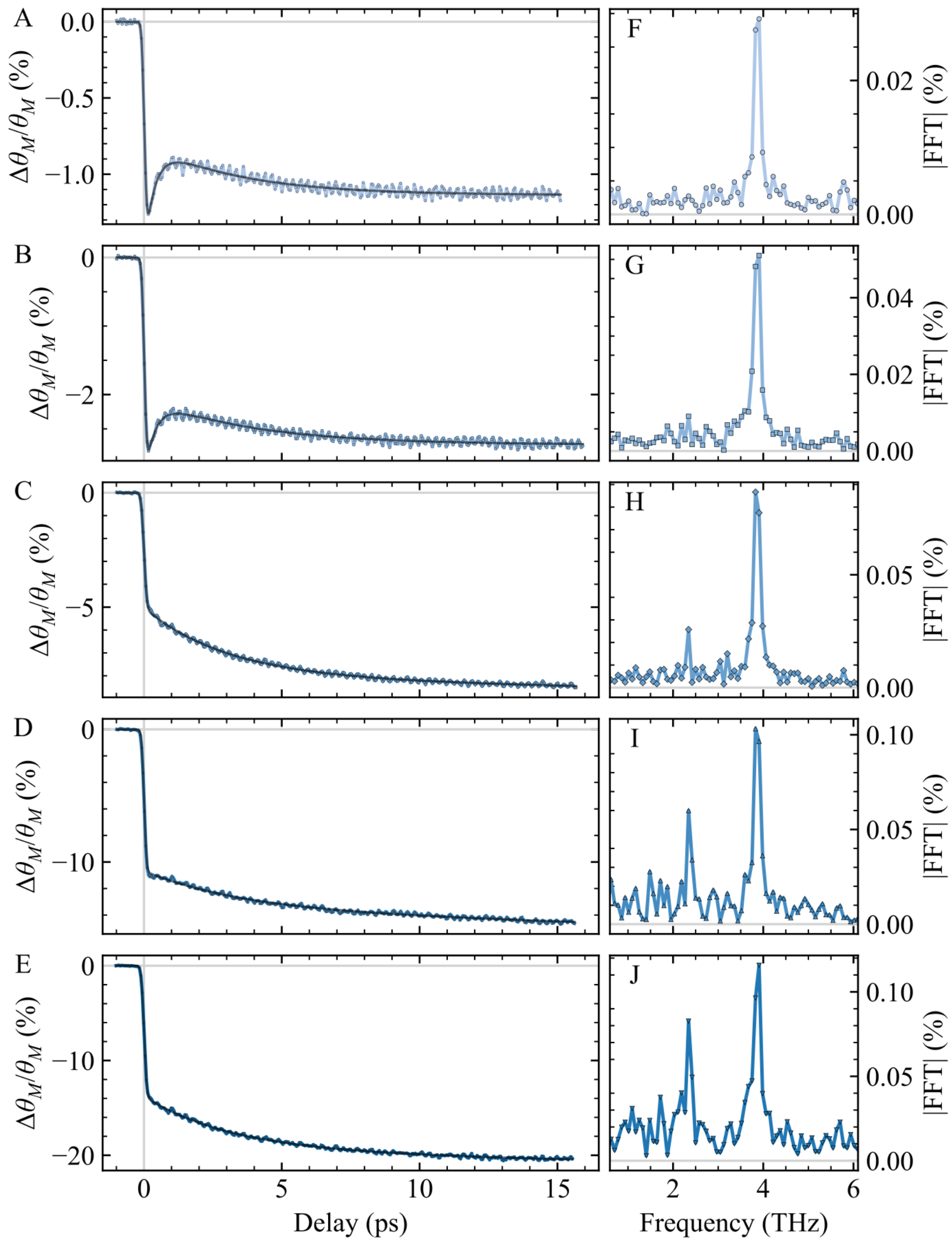

**Fig. S3b. Extended dataset for the B exciton, magnetic rotation.** Panels A-E show the measured data with the average temporal evolution fit for varying fluences (20, 50, 100, 190 and 215 μJ/cm$^2$, respectively). These measurements were recorded concurrently with those shown in Fig. S3a. Panels F-J show the fast Fourier transforms of the fit residuals. The fit parameters for the solid line are summarized in Table S1.

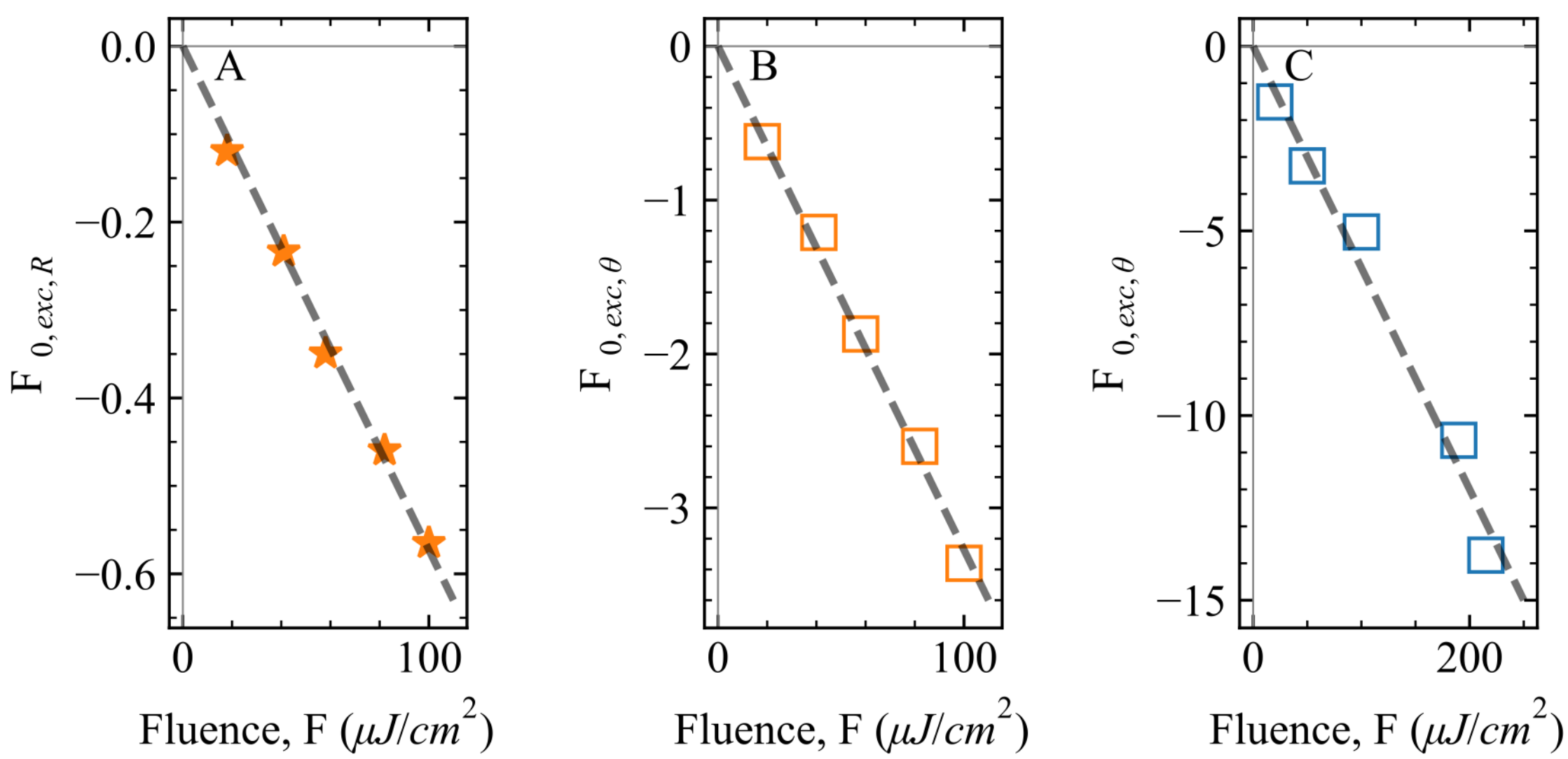

**Fig. S4. Selected parameters for initial response upon pumping of A and B excitons.** Panel A and B show the amplitude of the initial response for reflectivity and Kerr rotation upon pumping of exciton A. Panel C shows the same quantity for Kerr rotation upon pumping of exciton B.

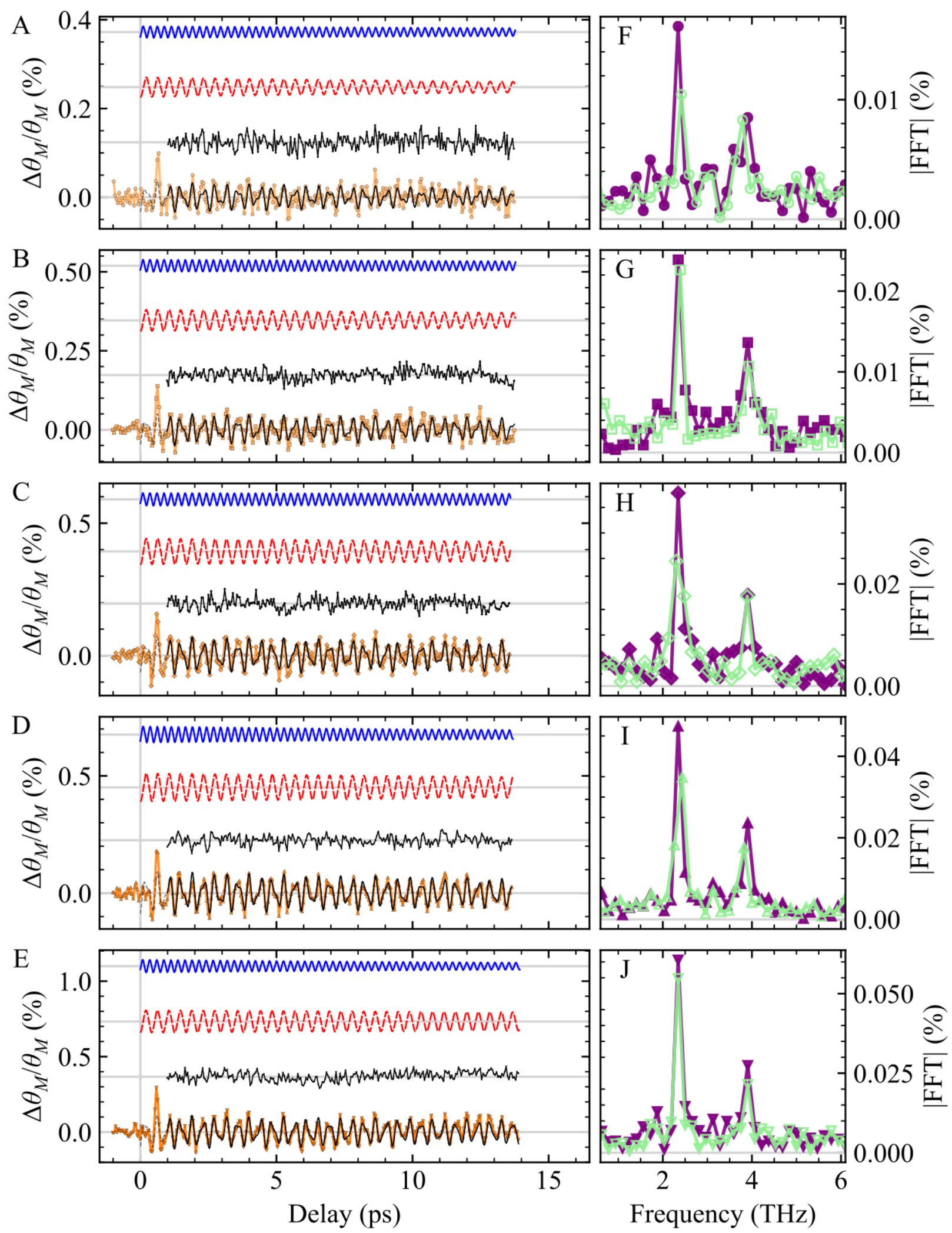

**Fig. S5. Fitting the oscillations of exciton A in Magnetic rotation.** Panels A-E show the extracted oscillation data with the two decaying sine modes fit for varying fluences (18, 41, 58, 82 and 100 µJ/cm$^2$, respectively) on the lowest horizontal line of the panels. One row above the data is the fit residual, above that the low frequency mode (dashed red line) and above that the high frequency mode (full line). Panels F-J show the fast Fourier transforms of the data for ranges 1-8 ps (purple lines with full symbols) and 8-15 ps (green line with empty symbols).

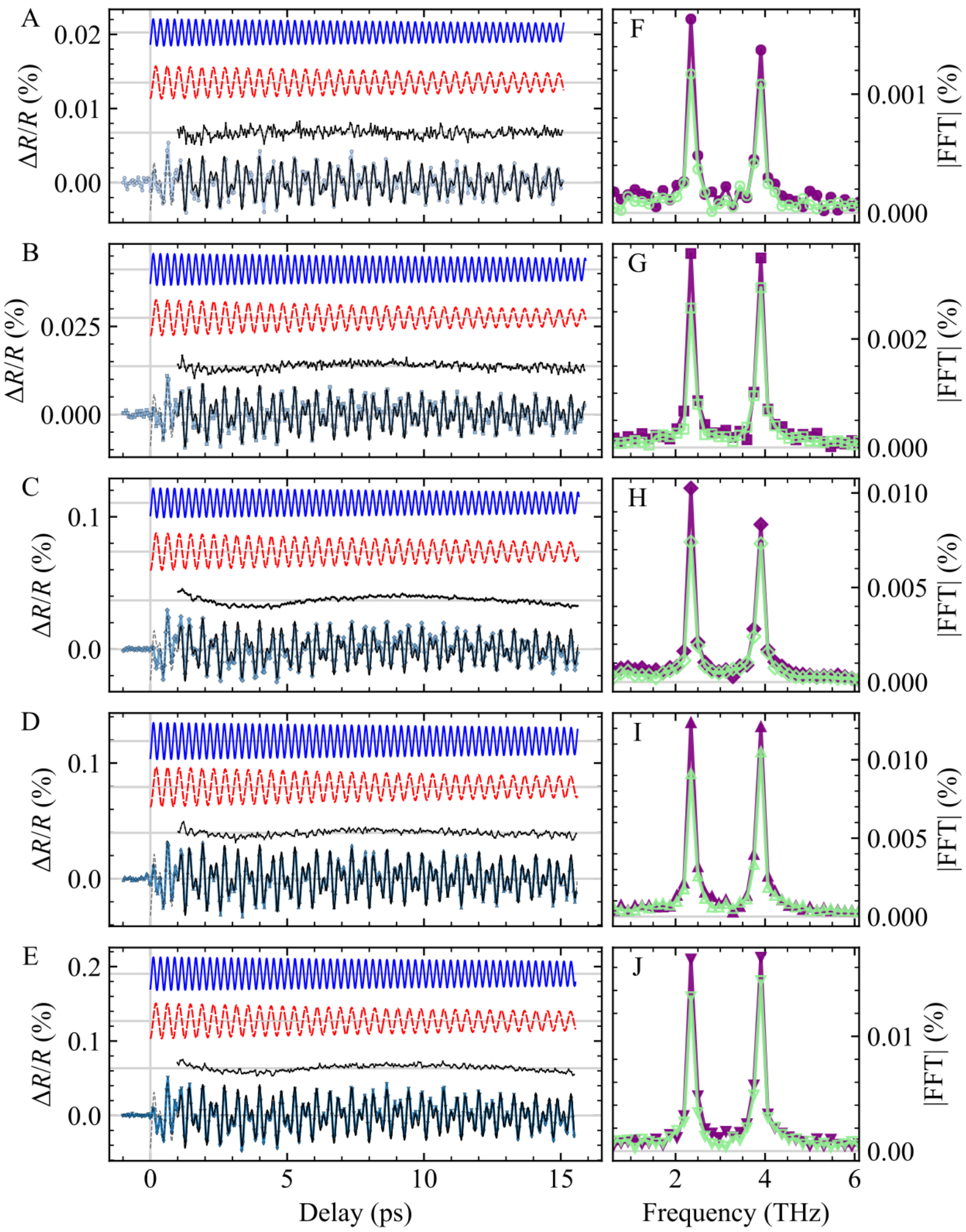

**Fig. S6a. Fitting the oscillations of exciton B in reflectivity.** Panels A-E show the extracted oscillation data with the two decaying sine modes fit for varying fluences (20, 50, 100, 190 and 215 μJ/cm$^2$, respectively) on the lowest horizontal line of the panels. One row above the data is the fit residual, above that the low frequency mode (dashed red line) and above that the high frequency mode (full line). Panels F-J show the fast Fourier transforms of the data for ranges 1-8 ps (purple lines with full symbols) and 8-15 ps (green line with empty symbols).

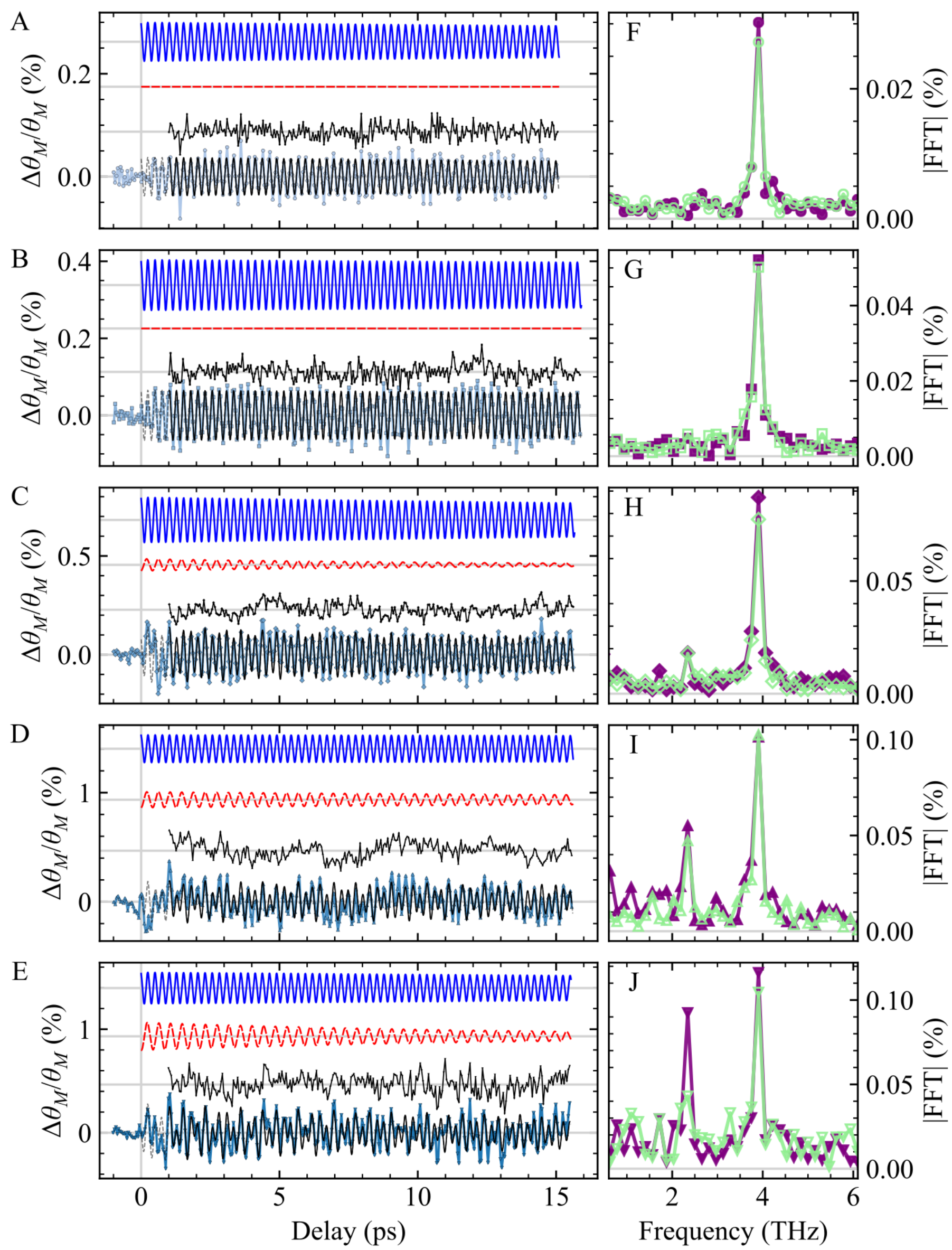

**Fig. S6b. Fitting the oscillations of exciton B in Kerr rotation.** Panels A-E show the extracted oscillation data with the two decaying sine modes fit for varying fluences (20, 50, 100, 190 and 215 µJ/cm$^2$, respectively) on the lowest horizontal line of the panels. One row above the data is the fit residual, above that the low frequency mode (dashed red line) and above that the high frequency mode (full line). Panels F-J show the fast Fourier transforms of the data for ranges 1-8 ps (purple lines with full symbols) and 8-15 ps (green line with empty symbols).

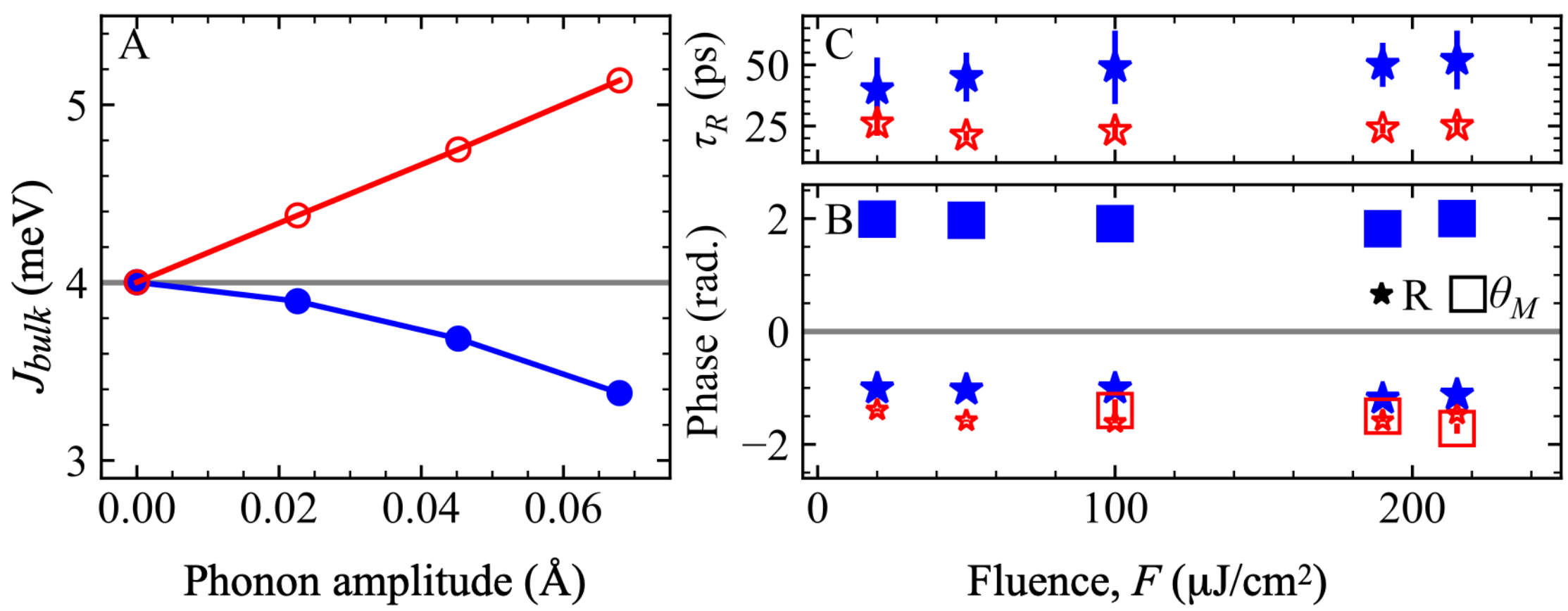

**Fig. S7. Phonon-driven spin precession.** (**A**) The results of frozen phonon calculations (see methods) show that the 3.9 THz (solid blue symbols) and 2.4 THz (open red symbols) phonon modes affect the isotropic magnetic exchange, $J^{iso}$, with opposite sign. (**B**) Phases of the phonon (stars) and spin oscillations (rectangles) for 3.9 THz (blue solid symbols) and 2.4 THz (red open symbols) summarized in Table S2 confirm the observation of this 180° phase shift following pumping of B excitons. (**C**) Damping times of the 2.4 THz (red open symbols) and 3.9 THz (blue solid symbols) phonon modes after pumping of B excitons from Table S3.

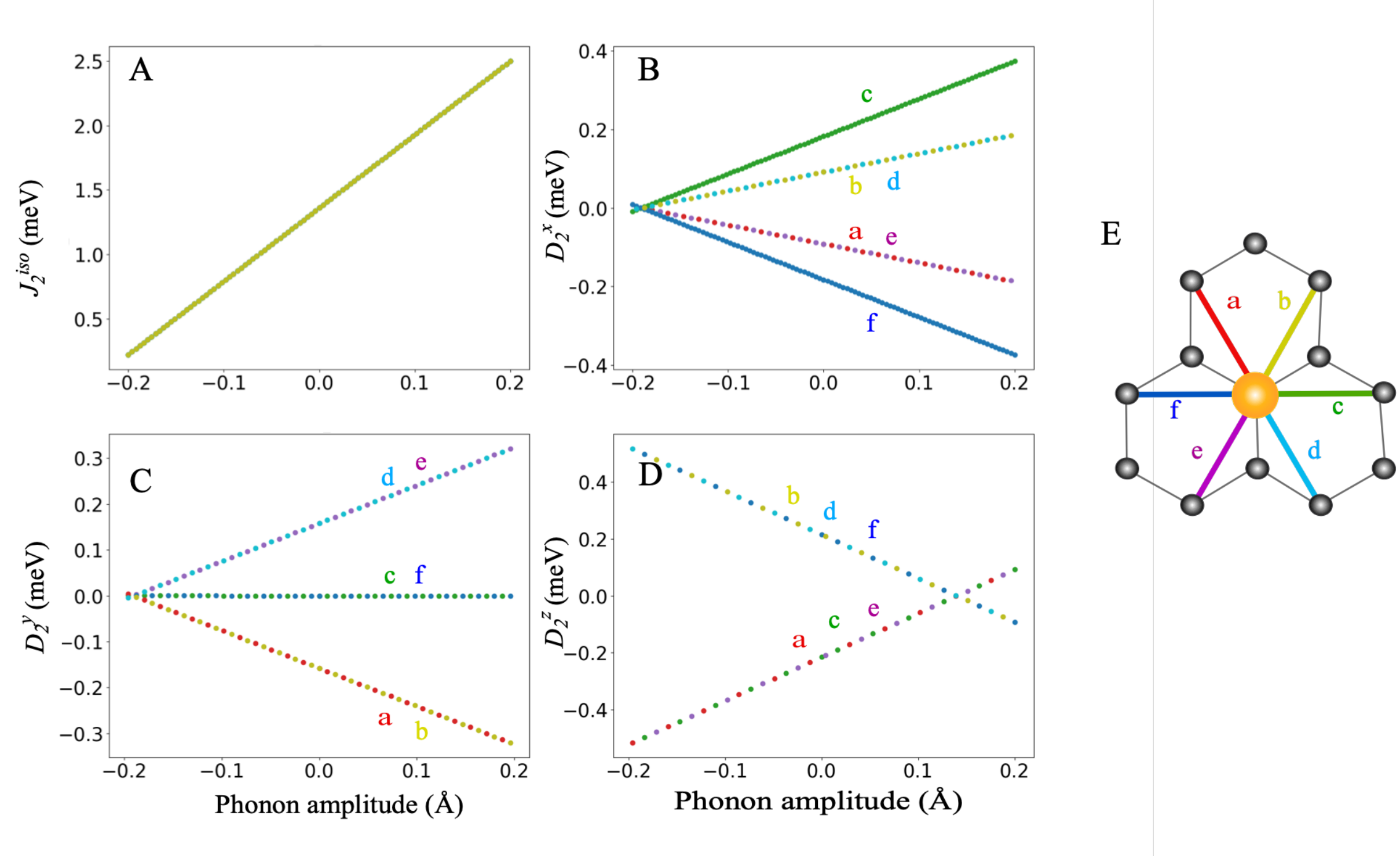

**Fig. S8. Effect of the 2.4 THz phonon mode on magnetic exchange interactions.** (**A**) Isotropic nearest-neighbor exchange, $J_2^{iso}$(see methods) vs. 2.4 THz phonon mode amplitude. (**B**) - (**D**) the three components $\alpha$ = x, y, z of the next-nearest neighbor chiral exchange, $D_2^{\alpha}$ (second term in eq. M1) for the exciton-Cr bonds indicated in panel (**E**).

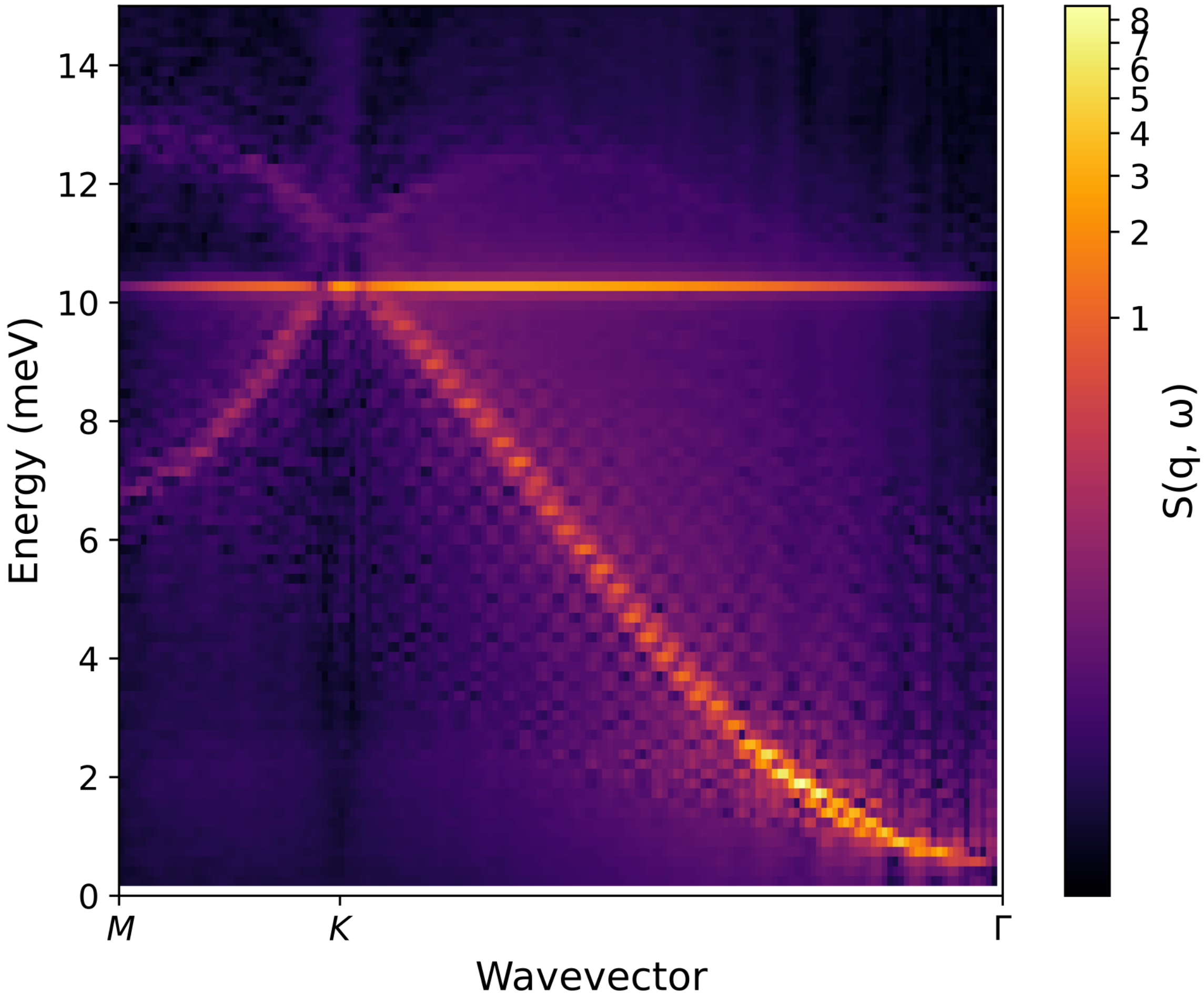

**Fig. S9. $CrI_3$ magnon dispersions.** Shown is the calculated spectral intensity distribution $S(q,\omega)$ while driving 2.4 THz phonons. The horizontal line is caused by the driven non-equilibrium magnons carrying orbital angular momentum while the other lines visible correspond to ground-state magnon dispersions.

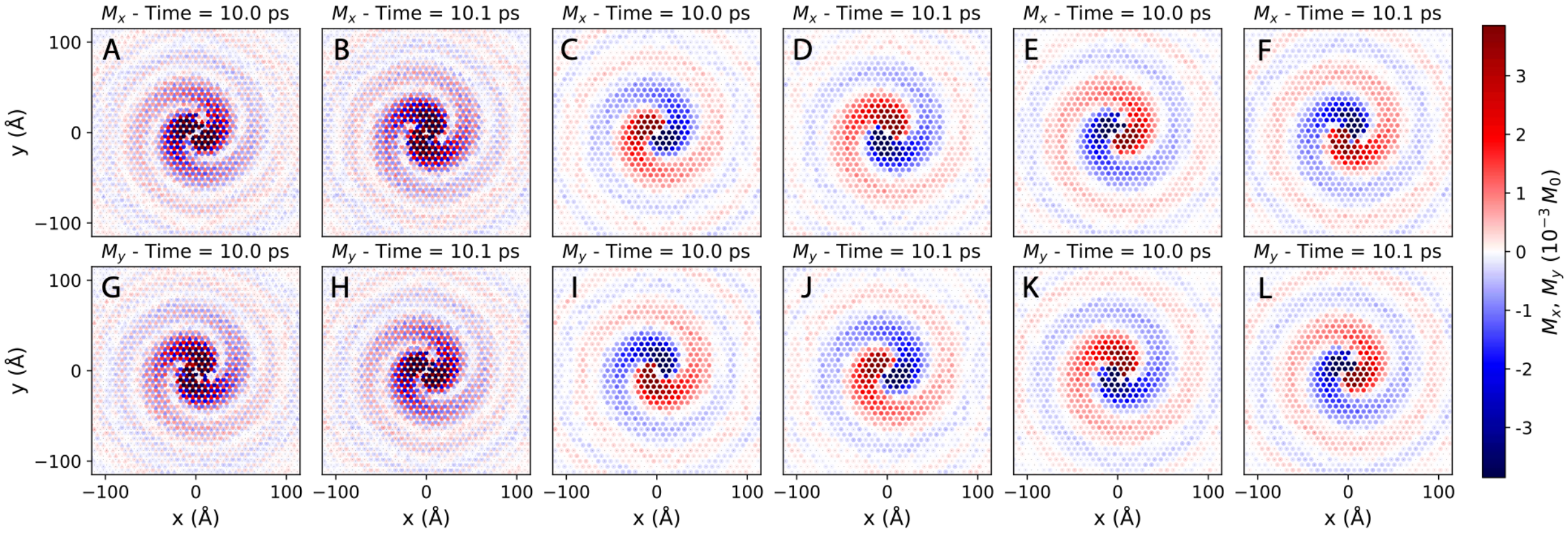

**Fig. S10. Exciton-generated spin motion.** Spin dynamics calculations (see methods) with phonon-spin coupling as in Fig. S8. (**A**, **B**, **G**, **H**) show the spin dynamics for both A and B magnetic sublattices. (**C**, **D, I, J**) show the spin dynamics for sublattice A. (**E**, **F**, **K**, **L**) show the spin dynamics for sublattice B. The $M_x$ (**A**, **C, E**) and $M_y$ (**G**, **I, K**) components are shown for 10.0 ps. (**B**, **D, F**) and (**H**, **J, L**) show the $M_x$ and $M_y$ components, respectively, for 10.1 ps. The time difference corresponds to one quarter of the atomic spin precession period of the 2.4 THz mode. Magnetic sublattices A and B correspond to the atomic moments connected to the central exciton via chiral and linear exchange interactions, respectively (see Fig. S8E).

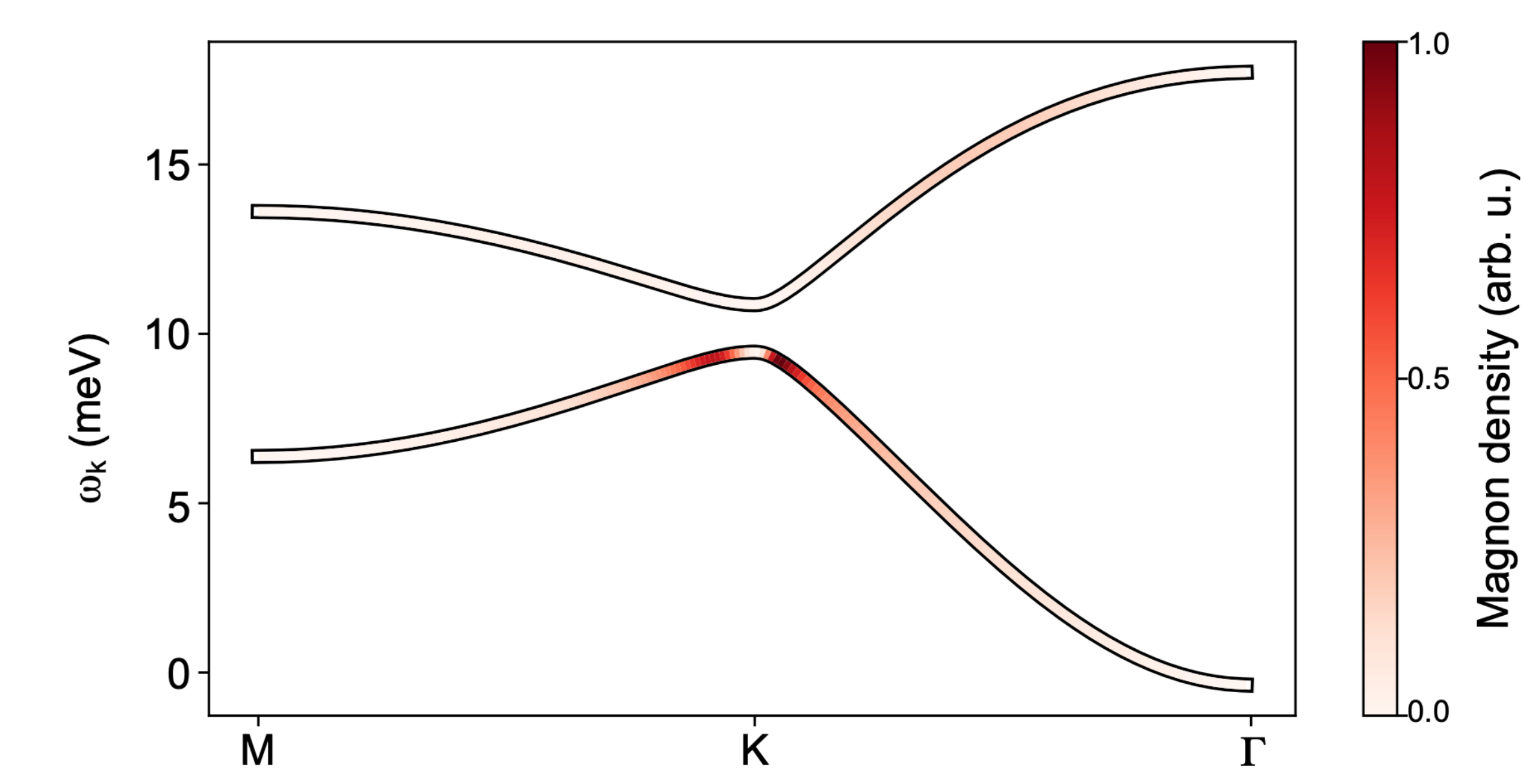

**Fig. S11: Magnon density on top of the dispersion relations.** Linear spin wave calculations of magnon dispersion following ref. *(7)*. The magnon density is calculated for resonant driving of the magnons with $\Omega = \omega_{K_1}^{\beta}$ as described in the methods section.

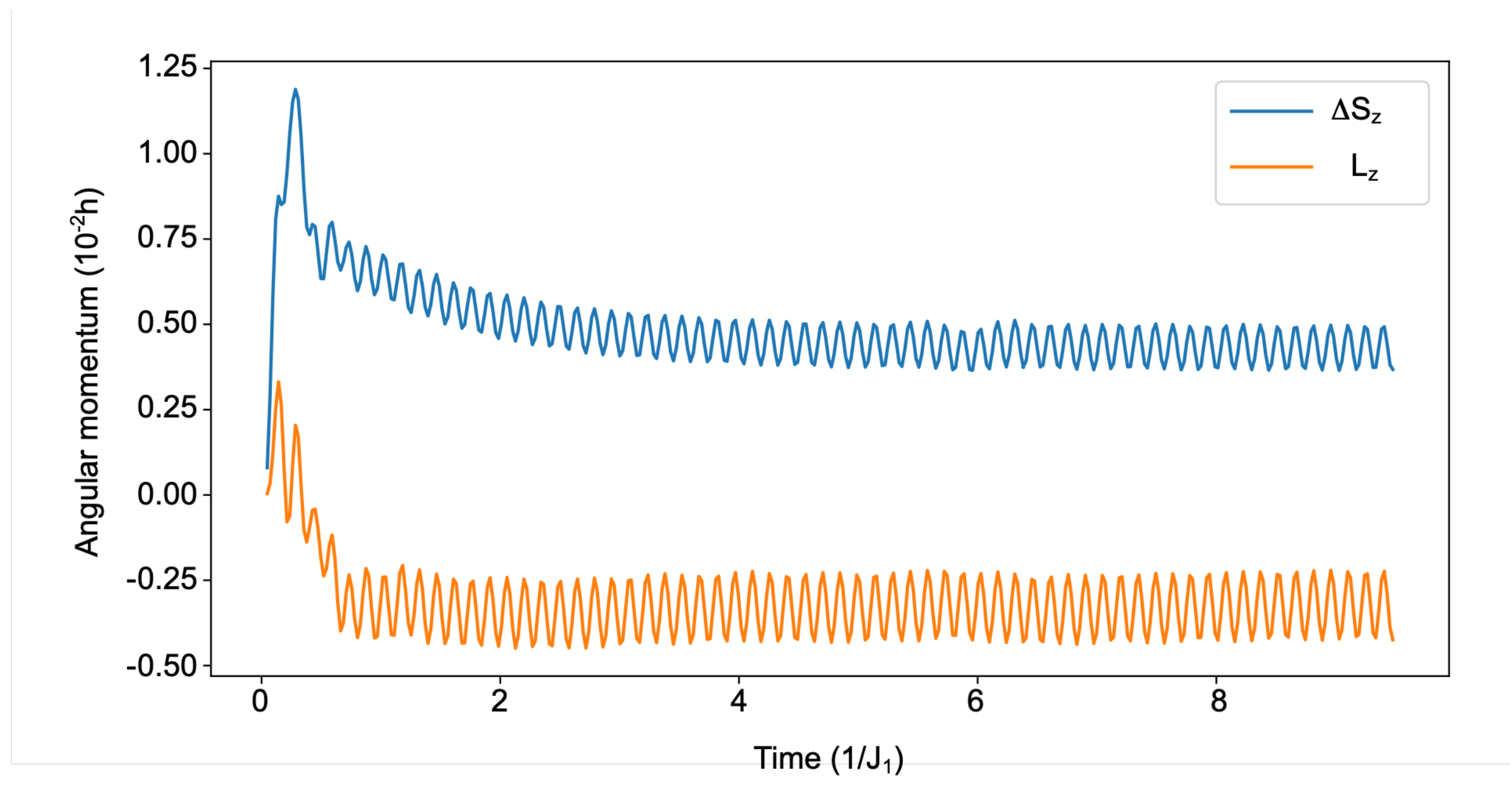

**Fig. S12: Change in spin and orbital angular momentum** calculated using linear spin wave theory (see methods) as a function of time for resonant driving of the magnon with $\Omega = \omega_{K_1}^{\beta}$. We use a damping of $\epsilon = 0.2$.

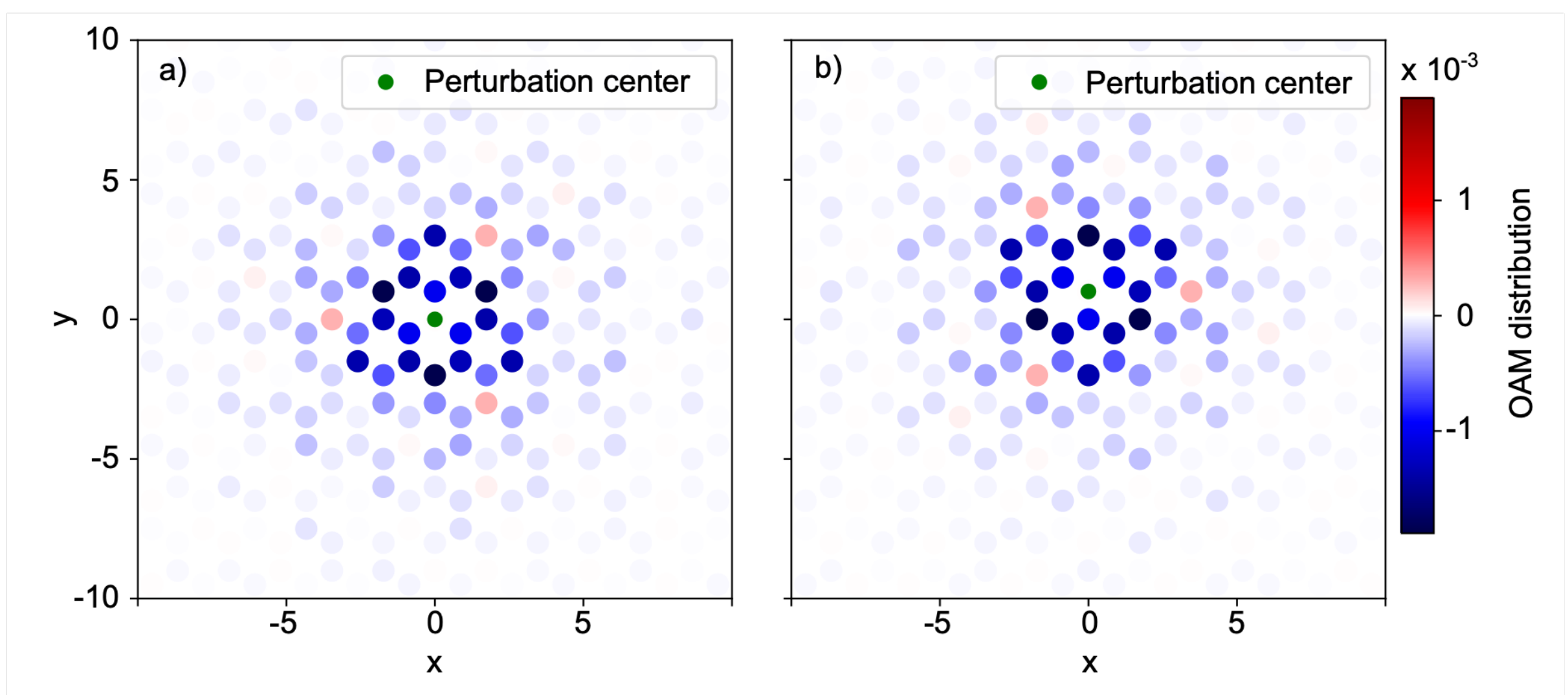

**Fig. S13. Orbital angular momentum distribution.** Calculations using linear spin wave theory (see methods) provide snapshots in a) and b) of the OAM distribution of $L_z$ with the driving exciton positioned at sublattice A and B, respectively.

| Type | Exciton | Fluence ($\mu$J/cm$^2$) | $\tau_{pulse}$ (fs) | F$_{0,\ exc}$ (%) | F$_{0,\ heat}$ (%) | $\tau_{heat}$ (ps) | F$_{0,\ decay}$ (%) | $\tau_{decay}$ (ps) |
|---|---|---|---|---|---|---|---|---|
| R | A | 18 | 192 ± 6 | -0.1193 ± 0.0007 | 0.0217 ± 0.0008 | 1.7 ± 0.1 | 0.24 ± 0.07 | 1000 * |
| | | 41 | 191 ± 3 | -0.2335 ± 0.0007 | 0.0575 ± 0.0009 | 2.02 ± 0.07 | 1.01 ± 0.08 | 1000 * |
| | | 58 | 197 ± 2 | -0.3494 ± 0.0008 | 0.1029 ± 0.0008 | 1.71 ± 0.03 | 2.64 ± 0.07 | 1000 * |
| | | 82 | 199 ± 2 | -0.4596 ± 0.0009 | 0.149 ± 0.001 | 2.12 ± 0.04 | 2.7 ± 0.1 | 1000 * |
| | | 100 | 205 ± 2 | -0.565 ± 0.001 | 0.183 ± 0.001 | 2.08 ± 0.03 | 3.54 ± 0.08 | 1000 * |
| $\theta_M$ | A | 18 | 210 ± 20 | -0.62 ± 0.01 | -0.078 ± 0.01 | 1.5 ± 0.4 | -4.5 ± 0.7 | 1000 * |
| | | 41 | 210 ± 10 | -1.21 ± 0.01 | -0.22 ± 0.01 | 1.5 ± 0.2 | -12.0 ± 1.0 | 1000 * |
| | | 58 | 220 ± 10 | -1.87 ± 0.02 | -0.5 ± 0.02 | 1.3 ± 0.1 | -32.0 ± 1.0 | 1000 * |
| | | 82 | 224 ± 8 | -2.6 ± 0.02 | -0.78 ± 0.02 | 1.17 ± 0.07 | -52.0 ± 1.0 | 1000 * |
| | | 100 | 228 ± 9 | -3.36 ± 0.03 | -0.91 ± 0.03 | 1.41 ± 0.08 | -61.0 ± 1.0 | 1000 * |
| R | B | 20 | 210 ± 10 | -0.058 ± 0.001 | 0.047 ± 0.001 | 0.62 ± 0.02 | -0.25 ± 0.03 | 1000 * |
| | | 50 | 200 ± 20 | -0.096 ± 0.002 | 0.087 ± 0.002 | 0.76 ± 0.03 | -0.38 ± 0.06 | 1000 * |
| | | 100 | 170 ± 20 | -0.209 ± 0.005 | 0.297 ± 0.005 | 1.3 ± 0.04 | 1.4 ± 0.2 | 1000 * |
| | | 190 | 200 ± 50 | -0.103 ± 0.007 | 0.264 ± 0.006 | 1.08 ± 0.05 | -1.0 ± 0.3 | 1000 * |
| | | 215 | 160 ± 60 | -0.116 ± 0.009 | 0.396 ± 0.008 | 1.18 ± 0.05 | -1.5 ± 0.4 | 1000 * |
| $\theta_M$ | B | 20 | 174 ± 7 | -1.51 ± 0.02 | -0.35 ± 0.02 | 3.3 ± 0.2 | 0.72 ± 0.03 | 0.37 ± 0.02 |
| | | 50 | 178 ± 6 | -3.24 ± 0.04 | -0.68 ± 0.03 | 3.9 ± 0.2 | 1.19 ± 0.04 | 0.37 ± 0.02 |
| | | 100 | 194 ± 6 | -5.03 ± 0.03 | -3.03 ± 0.05 | 3.1 ± 0.1 | -26.0 ± 4.0 | 1000 * |
| | | 190 | 167 ± 4 | -10.67 ± 0.04 | -2.9 ± 0.1 | 3.6 ± 0.2 | -63.0 ± 9.0 | 1000 * |
| | | 215 | 186 ± 4 | -13.78 ± 0.05 | -5.8 ± 0.1 | 3.2 ± 0.1 | -61.0 ± 9.0 | 1000 * |

**Table S1. Fit parameters for the average temporal evolution function.** Resulting parameters for the average temporal evolution fitting over the using equation SE1. The parameter $\tau_{\text{decay}}$ is kept fixed for the sets marked with (*) due to the later dynamics after limited fitting window here (approximately 15 ps).

| Type | Exciton | Fluence ($\mu J/cm^2$) | $G_{LF}$ (%) | $G_{HF}$ (%) | $\varphi_{LF}$ (rad) | $\varphi_{HF}$ (rad) | $\tau_{LF}$ (ps) | $\tau_{HF}$ (ps) |
|---|---|---|---|---|---|---|---|---|
| $\theta_M$ | A | 18 | 0.022 ± 0.004 | 0.013 ± 0.003 | -1.66 ± 0.08 | -1.00 ± 0.10 | 23 ± 12 | 38 ± 49 |
| | | 41 | 0.034 ± 0.004 | 0.019 ± 0.004 | -1.67 ± 0.05 | -0.73 ± 0.09 | 42 ± 25 | 47 ± 53 |
| | | 58 | 0.049 ± 0.004 | 0.024 ± 0.004 | -1.67 ± 0.04 | -0.65 ± 0.08 | 51 ± 29 | 114 ± 80 |
| | | 82 | 0.060 ± 0.004 | 0.035 ± 0.004 | -1.65 ± 0.03 | -0.92 ± 0.06 | 45 ± 18 | 24 ± 10 |
| | | 100 | 0.074 ± 0.006 | 0.042 ± 0.007 | -1.52 ± 0.04 | -0.51 ± 0.08 | 71 ± 52 | 24 ± 12 |
| R | B | 20 | 0.0022 ± 0.0001 | 0.0018 ± 0.0001 | -1.39 ± 0.03 | -1.01 ± 0.03 | 26 ± 5 | 40 ± 13 |
| | | 50 | 0.0051 ± 0.0002 | 0.0045 ± 0.0002 | -1.58 ± 0.02 | -1.03 ± 0.02 | 21 ± 2 | 45 ± 10 |
| | | 100 | 0.0142 ± 0.0007 | 0.0112 ± 0.0006 | -1.61 ± 0.02 | -1.01 ± 0.03 | 23 ± 3 | 49 ± 15 |
| | | 190 | 0.0174 ± 0.0005 | 0.016 ± 0.0005 | -1.58 ± 0.02 | -1.19 ± 0.01 | 24 ± 2 | 50 ± 9 |
| | | 215 | 0.024 ± 0.0009 | 0.0224 ± 0.0008 | -1.46 ± 0.02 | -1.12 ± 0.02 | 25 ± 3 | 52 ± 12 |
| $\theta_M$ | B | 20 | - | 0.038 ± 0.003 | - | 1.99 ± 0.03 | - | 75 ± 45 |
| | | 50 | - | 0.066 ± 0.003 | - | 1.97 ± 0.02 | - | 156 ± 141 |
| | | 100 | 0.031 ± 0.008 | 0.115 ± 0.006 | -1.40 ± 0.20 | 1.91 ± 0.03 | 12 ± 6 | 55 ± 18 |
| | | 190 | 0.07 ± 0.02 | 0.13 ± 0.01 | -1.50 ± 0.10 | 1.82 ± 0.05 | 33 ± 27 | 322 ± 296 |
| | | 215 | 0.13 ± 0.02 | 0.15 ± 0.02 | -1.72 ± 0.09 | 2.00 ± 0.05 | 14 ± 5 | 67 ± 58 |

**Table S2. Fit parameters for the two sine decaying modes.** The frequencies of $f_{LF}$ = 2.375 THz and $f_{HF}$= 3.865 THz in equation SE2 are used for the fits.

| Type | Fluence (µJ/cm$^2$) | $G_{LF}$ (%) | $G_{HF}$ (%) | $\varphi_{LF}$ (rad.) | $\varphi_{HF}$(rad.) | $\tau_{LF}$ (ps) | $\tau_{HF}$ (ps) | $\sigma$ (%) | $R^2$ |
|---|---|---|---|---|---|---|---|---|---|
| I | 20 | 0.0017 ± 0.0001 | 0.0015 ± 0.0001 | -1.38 ± 0.03 | -1.00 ± 0.03 | - | - | 0.0006058 | 0.8722 |
| | 50 | 0.0035 ± 0.0001 | 0.0038 ± 0.0001 | -1.58 ± 0.03 | -1.01 ± 0.02 | - | - | 0.0010986 | 0.9163 |
| | 100 | 0.0101 ± 0.0003 | 0.0095 ± 0.0003 | -1.61 ± 0.03 | -1.00 ± 0.03 | - | - | 0.0032582 | 0.9006 |
| | 190 | 0.0125 ± 0.0003 | 0.0136 ± 0.0003 | -1.57 ± 0.02 | -1.18 ± 0.02 | - | - | 0.0030079 | 0.9498 |
| | 215 | 0.0176 ± 0.0004 | 0.0192 ± 0.0004 | -1.46 ± 0.02 | -1.12 ± 0.02 | - | - | 0.0046996 | 0.9389 |
| II | 20 | 0.0021 ± 0.0001 | 0.0019 ± 0.0001 | -1.39 ± 0.03 | -1.01 ± 0.03 | 31 ± 5 | 31 ± 5 | 0.0005698 | 0.8869 |
| | 50 | 0.0047 ± 0.0002 | 0.0049 ± 0.0002 | -1.58 ± 0.02 | -1.04 ± 0.02 | 29 ± 3 | 29 ± 3 | 0.0009638 | 0.9355 |
| | 100 | 0.0131 ± 0.0005 | 0.0122 ± 0.0005 | -1.61 ± 0.02 | -1.01 ± 0.03 | 31 ± 4 | 31 ± 4 | 0.0029730 | 0.9173 |
| | 190 | 0.0160 ± 0.0004 | 0.0172 ± 0.0004 | -1.58 ± 0.02 | -1.19 ± 0.02 | 33 ± 3 | 33 ± 3 | 0.0025081 | 0.9651 |
| | 215 | 0.0222 ± 0.0007 | 0.0240 ± 0.0007 | -1.46 ± 0.02 | -1.12 ± 0.02 | 35 ± 4 | 35 ± 4 | 0.0041504 | 0.9523 |
| III | 20 | 0.0022 ± 0.0001 | 0.0018 ± 0.0001 | -1.39 ± 0.03 | -1.01 ± 0.03 | 26 ± 5 | 40 ± 13 | 0.0005684 | 0.8875 |
| | 50 | 0.0051 ± 0.0002 | 0.0045 ± 0.0002 | -1.58 ± 0.02 | -1.03 ± 0.02 | 21 ± 2 | 45 ± 10 | 0.0009427 | 0.9383 |
| | 100 | 0.0142 ± 0.0007 | 0.0112 ± 0.0006 | -1.61 ± 0.02 | -1.01 ± 0.03 | 23 ± 3 | 49 ± 15 | 0.0029381 | 0.9192 |
| | 190 | 0.0174 ± 0.0005 | 0.0160 ± 0.0005 | -1.58 ± 0.02 | -1.19 ± 0.01 | 24 ± 2 | 50 ± 9 | 0.0024406 | 0.9670 |
| | 215 | 0.0240 ± 0.0009 | 0.0224 ± 0.0008 | -1.46 ± 0.02 | -1.12 ± 0.02 | 25 ± 3 | 52 ± 12 | 0.0040762 | 0.9540 |

**Table S3. Comparison of three alternative models for coherent oscillations in reflectivity upon B-exciton pumping.** Shown are the fit parameters obtaind using equation SE2 for I no decay of both modes, II one identical decay for both modes, III two different decays for both modes. Quality of the fits is measured by the calculated residual standard deviation σ around the zero line and the $R^2$ parameter.

**Movie S1:** Enlarged view of the magnetization for the DMI (blue arrows) and J (red arrows) magnetic sublattices of the displayed monolayer slab of $CrI_3$ with periodic boundary conditions (see methods). A snapshot of the magnetization distribution at 100 fs is shown in Fig. 4A.

**Movie S2:** $M_x$ magnetization component of the displayed monolayer slab of $CrI_3$ with periodic boundary conditions (see methods).

**Movie S3:** Enlarged view of the $M_x$ magnetization component for the DMI magnetic sublattice of the displayed monolayer slab of $CrI_3$ with periodic boundary conditions (see methods). A snapshot at 10.0 ps is displayed in Fig. 4B.